\documentclass[final]{siamltex}

\usepackage{amssymb}
\usepackage{epsf}

\renewcommand{\leq}{\leqslant}  

\def\condmath#1{\leavevmode\ifmmode #1 \else $#1$ \fi}
\newcommand{\calf}{\ensuremath{\mathcal{F}}}

\newcommand{\call}{\ensuremath{\mathcal{L}}}
\newcommand{\calk}{\ensuremath{\mathcal{K}}}
\newcommand{\calg}{\ensuremath{\mathcal{G}}}
\newcommand{\calp}{\ensuremath{\mathcal{P}}}

\newcommand{\ie}{{i.e.}}
\newcommand{\ZZ}{\condmath{Z \! \! \! Z}}
\renewcommand{\Re}{\condmath{I \! \! R}}

\newtheorem{rem}[theorem]{Remark}
\newenvironment{Rem}{\begin{rem}\begin{em}}{\end{em}\end{rem}}

\newcommand{\ch}{\mbox{CH}}
\newcommand{\n}{\ensuremath {\{1,\ldots,n\}}}
\newcommand{\Ae}{\ensuremath{{\cal A}_e}}
\newcommand{\ko}{\ensuremath {k_{i_0}}}

\newcommand{\si}{\ensuremath{s_i}}
\newcommand{\sj}{\ensuremath{s_j}}
\newcommand{\so}{\ensuremath{s_{i_0}}}

\newcommand{\A}{\ensuremath{{\cal A}}}

\newcommand{\E}{\ensuremath {{\cal E}}}
\renewcommand{\S}{\ensuremath{{\cal S}}}
\renewcommand{\L}{\ensuremath{{\cal L}}}

\newcommand{\F}{\ensuremath{{\cal F}}}
\newcommand{\df}{\ensuremath{\partial(\F)}}
\newcommand{\Fp}{\ensuremath{{\cal F}_e}}

\newcommand{\dfp}{\ensuremath{\partial(\Fp)}}

\newcommand{\ci}{\ensuremath {C_i}}
\newcommand{\co}{\ensuremath {C_{i_0}}}
\newcommand{\cso}{\ensuremath {C_{s_{i_0}}}}
\newcommand{\cei}{\ensuremath {C_{e_i}}}

\newcommand{\ceio}{\ensuremath {C_{e_{i_0}}}}

\newcommand{\Ci}{\ensuremath {{\cal C}_i}}

\newcommand{\Co}{\ensuremath {{\cal C}_{i_0}}}
\newcommand{\Cso}{\ensuremath {{\cal C}_{s_{i_0}}}}

\renewcommand{\H}{\ensuremath{{\cal H}}}
\newcommand{\Hi}{\ensuremath{{\cal H}_i}}

\newcommand{\Hsi}{\ensuremath{{\cal H}_{s_i}}}

\newcommand{\Hdi}{\ensuremath{{\cal H}_{D_i}}}

\newcommand{\Hei}{\ensuremath{{\cal H}_{e_i}}}

\newcommand{\Z}{\ensuremath{{\cal Z}}}
\newcommand{\Zi}{\ensuremath{{\cal Z}_i}}

\newcommand{\Zsi}{\ensuremath{{\cal Z}_{s_i}}}

\newcommand{\Zdi}{\ensuremath{{\cal Z}_{D_i}}}

\newcommand{\Zei}{\ensuremath{{\cal Z}_{e_i}}}

\newcommand{\Ysi}{\ensuremath{{\cal Y}_{s_i}}}
\newcommand{\Ysip}{\ensuremath{{\cal Y}_{s'_i}}}
\newcommand{\Ydi}{\ensuremath{{\cal Y}_{D_i}}}
\newcommand{\Yei}{\ensuremath{{\cal Y}_{e_i}}}

\newcommand{\ei}{\ensuremath{e_i}}

\newcommand{\tti}{\ensuremath {{\cal T}_i}}
\newcommand{\rri}{\ensuremath {{\cal R}_i}}
\newcommand{\rrip}{\ensuremath {{\cal R}^+_i}}

\newcommand{\Om}{\ensuremath {\Omega}}
\newcommand{\Ga}{\ensuremath {\Gamma}}

\newcommand{\placeipe}[2]{ \begin{figure} \begin{center}
   \def\IPEfile{#1.ipe}\input{#1.ipe} \end{center} \caption{#2 \label{#1}}
   \end{figure}}

\title{Motion Planning of Legged Robots\thanks{
Part of these results have been presented in conferences
\cite{prisme-3214i,prisme-3214i2}.}}

\author{Jean-Daniel Boissonnat\footnotemark[2] \and Olivier Devillers\footnotemark[2] \and Sylvain Lazard\footnotemark[3]
}

\begin{document}

\maketitle

\renewcommand{\thefootnote}{\fnsymbol{footnote}}
\footnotetext[2]{INRIA Sophia-Antipolis, BP 93, 06902 Sophia Antipolis Cedex, France. \\
E-mail: firstname.name@sophia.inria.fr. http://www-sop.inria.fr/prisme/prisme\_eng.html.}
\footnotetext[3]{INRIA Lorraine, 615 rue du jardin botanique, B.P. 101, 54602
  Villers-les-Nancy Cedex,
  France. E-mail: lazard@loria.fr. http://www.loria.fr/~lazard/. Most of this work was
  done while this author was at INRIA Sophia-Antipolis.}
\renewcommand{\thefootnote}{\arabic{footnote}}

\begin{abstract}
We study the problem of computing the free space \F\ of a simple
legged robot called the spider robot. The body of this robot is a
single point and the legs are attached to the body.  The robot is
subject to two constraints: each leg has a maximal extension $R$
(accessibility constraint) and the body of the robot must lie above
the convex hull of its feet (stability constraint).  Moreover, the
robot can only put its feet on some regions, called the foothold
regions.  The free space $\calf$ is the set of positions of the body
of the robot such that there exists a set of accessible footholds for
which the robot is stable.  We present an efficient algorithm that
computes \F\ in $O(n^2\log n)$ time using $O(n^2\alpha(n))$ space 
for $n$ discrete point footholds
where $\alpha(n)$ is an extremely slowly growing function
($\alpha(n)\leq 3$ for any practical value of $n$).  We also present
an algorithm for computing \F\ when the foothold regions are pairwise
disjoint polygons with $n$ edges in total.  This algorithm computes
\F\ in $O(n^2\alpha_8(n)\log n)$ time using $O(n^2\alpha_8(n))$ space
($\alpha_8(n)$ is also an extremely slowly growing function).  These
results are close to optimal since $\Omega(n^2)$ is a lower bound for
the size of~$\F$.
\end{abstract}



\pagestyle{myheadings}
\thispagestyle{plain}
\markboth{J.-D. BOISSONNAT, O. DEVILLERS, AND S. LAZARD}{MOTION PLANNING OF LEGGED ROBOTS}


\section{Introduction}

Although legged robots have already been studied in
robotics \cite{legloc2,legloc}, only a very few papers consider the
motion planning problem amidst
obstacles~\cite{Hirose,Hirose2,bddp-mplrs-95}.
In~\cite{Hirose,Hirose2} some heuristic approaches are described
while, in~\cite{bddp-mplrs-95} efficient and provably correct
geometric algorithms are described for a restricted type of legged
robots, the so-called spider robots to be defined precisely below, and
for finite sets of point footholds.

A {\em legged robot} consists of a body with legs.  Each leg has one
end attached to the body and the other end (called the foot) that can
lie on the ground (or move in space between two positions on the
ground).  Compared to the classic piano movers problem, legged robots
introduce new types of constraints. We assume that the environment
consists of regions in the plane, called {\em foothold regions}, where
the robot can safely put its feet.  A {\em foothold} is a point in a
foothold region.  The legged robot must satisfy two different
constraints: the accessibility and the stability constraints.  A
foothold is said to be {\em accessible} from a {\em placement}
(position of the body of the robot) if it can be reached by a leg of
the robot.  A placement is called {\em stable} if there exist
accessible footholds and if the center of mass of the robot lies above
the convex hull of these accessible footholds.  The set of stable
placements is clearly relevant for planning the motion of a legged
robot: we call this set {\em the free space} of the legged robot. Note
that a legged robot has at least four legs, three legs ensure the
stability of a placement and a fourth leg permits the motion of the
robot.

 \begin{figure}[t] \begin{center}
  \unitlength 1cm \input{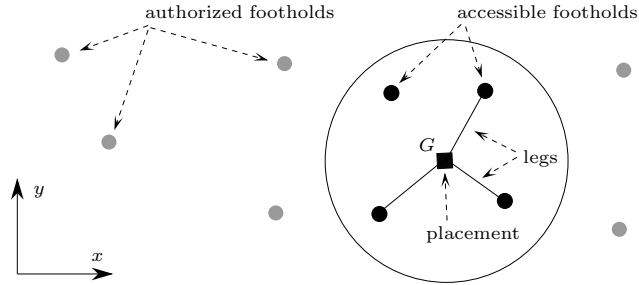} \end{center} \caption{The spider robot.}
  \label{Spider-plan} \end{figure}

A first simple instance of a legged robot is the {\em spider robot}
(see Figure~\ref{Spider-plan}). The spider robot was inspired by
 Ambler, developed at Carnegie Mellon University~\cite{bw}.  The
body of the spider robot is a single point in the Euclidean plane and
all its legs are attached to the body. The legs are retractable and
their lengths may vary between $0$ and a constant $R$. We also assume
that the center of mass of the robot is its body. It follows that a
placement is stable if the body of the robot lies above the convex
hull of the accessible footholds.

The constraint that the body of the spider robot lies in the plane
(instead of in 3D) is not really restrictive. Indeed, consider a
legged robot for which that constraint is relaxed. Then, if a
placement $(x,y,z)$ of such a legged robot is stable then, any
placement $(x,y,z')$, $0\leq z'\leq z$ is also stable.  Reciprocally, it
can be shown that if $(x,y)$ is in the interior of the free space of
the spider robot, then there exists $z>0$ such that $(x,y,z)$ is a
stable placement of the corresponding legged robot.

The problem of planning the motion of a spider robot 
has already been studied by Boissonnat et
al. \cite{bddp-mplrs-95}. However, their method assumes that the set of
footholds is a finite set of points and cannot be generalized to more
complex environments.
This paper proposes a new method for computing the
free space of a spider robot in the presence of polygonal foothold
regions. This method is based on a transformation between this problem
and the problem of moving a half-disk amidst obstacles.  Our method
requires the computation of some parts of the free space of the
half-disk.  These computations are rather
technical and complicated.  Consequently, for the sake of clarity, we
first present  our algorithm for the simple case of 
discrete footholds, then we show how it can be generalized to
the case of polygonal foothold regions.

Once the free space of the spider robot has been computed, it can be
used to find trajectories and sequences of legs assignments allowing
the robot to move from one point to another.  Indeed, once the free
space is known, a trajectory of the body can be found in the free
space. Then, a sequence of legs assignments can be computed as
follows (see~\cite{bddp-mplrs-95} for details).  Given an initial
legs assignment, the body of the robot moves along its trajectory
until it crosses the convex hull of its (three) feet that are on the ground or
one leg reaches its maximal extension. Then, a suitable foothold is
found for the fourth leg and one leg leaves its foothold.

The paper is organized as follows: some notations and results
of~\cite{bddp-mplrs-95} are recalled in the next
section. Section~\ref{fsrthdr} shows the transformation between the
spider robot problem and the half-disk problem. We present in
Section~\ref{Computation_of_F}
our algorithm for computing
 the free space of a spider robot 
for a discrete set of footholds.
Section~\ref{polygonal-foothold-regions} shows how to extend the
algorithm  to polygonal foothold regions.

\section{Notations and previous results}
\label{Notations_and_previous_results}

In Sections 2, 3 and 4, \S\ denotes a discrete set of
distinct  footholds $\{s_1,\ldots,s_n\}$ in the Euclidean
plane (\S\ will denote in Section~5 a set of disjoint polygonal
regions). Point $G$ denotes the body of the robot (in the
same plane) and $[0,R]$ is the length range of each leg.
The free space \F\ is the set of all stable placements of $G$.  A placement
is said to be at the {\em limit of stability} if it lies on the
boundary of the convex hull of its accessible 
footholds.  Notice that \F\ is a closed set and contains the
placements at the limit of stability.

Let $C_i$ denote the circle of radius $R$ centered at $\si$. \A\ is
the arrangement of the circles $C_i$ for $1\leq i\leq n$, \ie, the
subdivision of the plane induced by the circles. This arrangement
plays an important role in our problem and we will express the
complexity results in term of $|\A|$, the size of $\A$.  In the
worst-case, $|\A|=\Theta(n^2)$ but if $k$ denotes the maximum number 
of disks that can cover a point of the plane, among the disks 
of radius $R$ centered at the $s_i$,   it can be
shown that $|\A|= O(kn)$~\cite{s-ksacs-91}. Clearly $k$ is not larger
than $n$ and in case of sparse footholds, $|\A|$ may be linearly
related to the number of footholds.

For any set $\E$, let $\partial(\E)$ denote its boundary, $\ch(\E)$ its
convex hull, $int(\E)$ its relative interior\footnote{The relative
interior of a set $\E$ in a space $E$ is the interior of $\E$ in the
space $\E$ for the topology induced by $E$. For example, the relative
interior of a closed line segment in ${\Re}^3$ is the line segment
without its endpoints, though its interior in
${\Re}^3$ is empty.}, $clos(\E)$ its closure, and $compl(\E)$ its
complementary set.  Let $S^1$ denote the set of angles $\Re/2\pi\ZZ$.
We denote by $x=y[p]$ the equality of $x$ and $y$ modulo $p$.  We say
in the sequel that two objects {\em properly intersect} if and only if
their relative interiors intersect.

\begin{figure}[ht] \begin{center}
  \unitlength 1cm \input{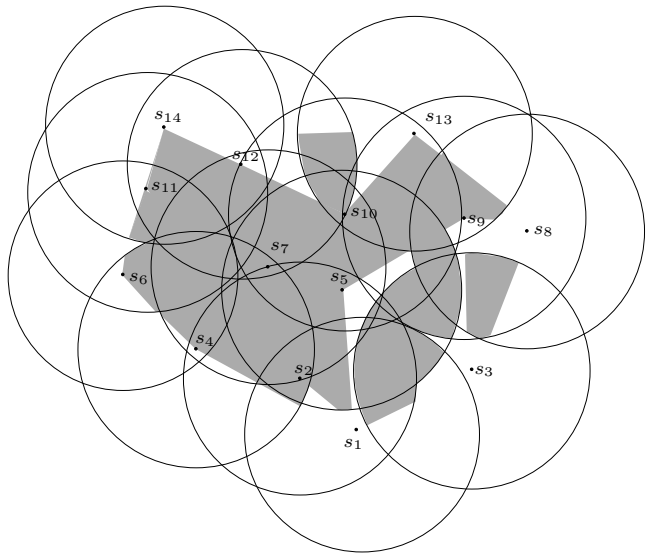} \end{center} \caption{An example of the free space of a spider
robot.}
  \label{exemple-F-2} \end{figure}

The algorithm described in~\cite{bddp-mplrs-95} is based on the
following observation: for $G$ in a cell \Ga\ of $\A$, the set of
footholds that can be reached by the robot is fixed; the portion of
\Ga\ that belongs to \F\ is exactly the intersection of \Ga\ with the
convex hull of the footholds that can be reached from $\Ga$.  Therefore,
the edges of \df\ are either circular arcs belonging to \A\ or
portions of line segments joining two footholds. Moreover, a vertex of
\df\ incident to two straight  edges is a foothold (see
Figure~\ref{exemple-F-2}). The complexity of \F\ has been proved to be
$|\F|=\Theta(|\A|)$~\cite{bddp-mplrs-95}.

The algorithm presented in~\cite{bddp-mplrs-95} computes the free
space \F\ in $O(|\A|\log n)$ time. It uses sophisticated data
structures allowing the off-line maintenance of convex hulls.

The algorithm described in this paper has the same time complexity,
uses simple data structures and can be extended to the case where the
set \S\ of footholds is a set of polygonal regions and not simply a
set of points. For simplicity, we consider first the case of point
footholds and postpone the discussion on polygonal foothold regions to
Section~\ref{polygonal-foothold-regions}.

\subsection*{General position assumption}

To simplify the presentation of this paper, we make the following
general position assumptions. All these hypotheses can be removed by a
careful analysis.  Recall that we consider here that the set of
footholds is discrete.

No two footholds lie at distance exactly $R$ or $2R$.  Among the
circles $C_1,\ldots,C_n$ and the line segments joining two footholds,
the intersection between three circles or, two circles and a line
segment or, one circle and two line segments, is empty.

\section{From spider robots to half-disk robots}
\label{fsrthdr}

In this section, we establish the connection between the free space of
the spider robot and the free space of a half-disk robot moving by
translation and rotation amidst $n$ point obstacles.

\begin{figure}[bht] \begin{center}
  \unitlength 1cm \input{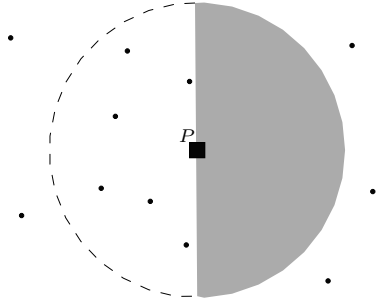} \end{center} \caption{A placement which is not stable.}
  \label{Spider-To-HD} \end{figure}

\begin{theorem}
\label{thFL1} 
The spider robot does not admit a stable placement at point $P$ if and
only if there exists a half-disk (of radius $R$) centered at $P$ that
does not contain any foothold of \S\ (see Figure~\ref{Spider-To-HD}).
\end{theorem}
\proof
Let $\cal R$ be the set of all the footholds that are reachable from
placement $P$.  By definition, $P$ is not stable if and only if the
convex hull of $\cal R$ does not contain $P$ (see
Figure~\ref{Spider-To-HD}).  That is equivalent to say that there
exists an open half-plane through $P$ containing $\cal R$, or that
there exists a closed half-disk of radius $R$ centered at $P$ which
does not contain any foothold.  
\endproof

 \begin{figure}[th] \begin{center}
  \unitlength 1cm \input{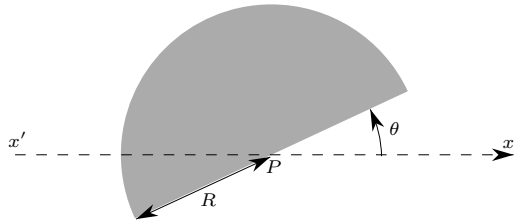} \end{center} \caption{$HD(P,\theta)$.}
  \label{def-DD} \end{figure}

\begin{definition}
\label{def-HD}
Let $HD(P,\theta)$ be the half-disk of radius $R$ centered at $P$ (see
Figure~\ref{def-DD}) defined by:
\[ 
\left\{ \begin{array}{l} 
(x-x_P)^2 + (y-y_P)^2 \leq R^2 \\ (x-x_P) \sin \theta - (y-y_P) \cos
\theta \leq 0
\end{array} \right. 
\] 
\end{definition}

\begin{definition}
\label{def-Hi} 
$\forall s_i\in\S \ (1\leq i\leq n)$ let:
\[\Hi = \{(P,\theta)\in\Re^2\times S^1\,|\ P\in HD(s_i,\theta)\},\]
\[\H=\bigcup_{i =1}^n\Hi,\]
\[\Ci =C_i\times S^1.\]
\Hi\ will be called the helicoidal volume centered at \si\ 
(see Figure~\ref{Hi}).
\end{definition}

\begin{figure}[ht]
\setlength{\epsfxsize}{8cm}
\setlength{\epsfysize}{12cm}
\centerline{\epsfbox{Youplaoup.epsi}}
\caption{Helicoidal volume $\Hi$.}
\label{Hi}
\end{figure}

Notice the typographical distinction between the circle $C_i$ defined
in ${\Re}^2$ and the torus \Ci\ defined in ${\Re}^2\times S^1$.  For
convenience, we will often identify $S^1$ and the interval $[0,2\pi]$
of $\Re$. This allows us to draw objects of ${\Re}^2\times S^1$ in
${\Re}^3$ and to speak of the $\theta$-axis.  $\Pi_{\theta_0}$ denotes
the ``plane'' $\{(P,\theta)\in\Re^2\times S^1\,|\ \theta =\theta_0\}$.

\begin{definition}
\label{def-space-L}
The free space \L\ of a half-disk robot moving by
translation and rotation amidst the set of obstacles \S\ is the set of
$(P,\theta)\in{\Re}^2\times S^1$ such that the half-disk
$HD(P,\theta+\pi)$ does not intersect  \S.
\end{definition}

\begin{proposition}
\label{propL} 
$\L=compl(\H)$.
\end{proposition}
\proof  
$\forall\theta\in S^1$, the set $\L\cap \Pi_{\theta}$ is the free
space of the half-disk $HD(P,\theta+\pi)$ moving by translation only,
amidst the obstacle $s_1,\ldots,s_n$.  Since the set of points $P$
such that $HD(P,\theta+\pi)$ contains a $s_i$ is $HD(\si,\theta)$,
$\L\cap \Pi_{\theta}$ is the complementary set of the union of the
$HD(\si,\theta)$. Thus, $\L$ is the complementary set of the union of the
$\Hi$, that is $\H$.
\endproof 

Let $p_{/\!/\theta}$ denote the mapping (called ``orthogonal projection''):
${\Re}^2\times S^1 \longrightarrow {\Re}^2, (P,\theta) \mapsto P$.
\begin{theorem}
\label{thFHi} 
$\F=compl(p_{/\!/\theta}(compl(\H)))$
\end{theorem}
\proof 
By definition of $\L$, $p_{/\!/\theta}(\L)$ is the set of points
$P\in{\Re}^2$ such that there exists an angle $\theta\in S^1$ such
that the half-disk $HD(P,\theta)$ does not intersect $\S$.  By
Theorem~\ref{thFL1}, it is equivalent to say that there exists
$\theta\in S^1$ such that $HD(P,\theta)$ does not intersect $\S$, or
that $P$ is not a stable placement of the spider robot.  Thus,
$p_{/\!/\theta}(\L)$ is the set of points $P$ where the robot does not
admit a stable placement, i.e., $\F=compl(p_{/\!/\theta}(\L))$.  
The result then follows from Proposition~\ref{propL}.
\endproof 

\begin{Rem}
\label{RemFHi} 
$compl(p_{/\!/\theta}(compl(\H)))\times S^1$ is the largest ``cylinder''
included in $\H$, whose axis is parallel to the $\theta$-axis (in grey in 
Figure~\ref{CpC}).  The basis of this cylinder is $\F$.
\end{Rem}

 \begin{figure}[ht] \begin{center}
  \unitlength .8cm \input{CpC.ltex} \end{center} \caption{\protect$compl(p_{/\!/\theta}(compl(\E)))$.}
  \label{CpC} \end{figure}

\begin{Rem}
\label{RemGeneralize}
The results of this section do not depend on the fact that the
footholds are discrete points. For more general foothold regions, we
simply need to replace the helicoidal volumes by their analog.  This
will be done in Section \ref{polygonal-foothold-regions}.
\end{Rem}

\section{Computation of \F}
\label{Computation_of_F}

In this section, we  propose an algorithm
for computing \F\ based on Theorem \ref{thFHi}.

A first attempt to use Theorem \ref{thFHi} may consist in 
computing $\L=compl(\H)$ and projecting it onto  
the horizontal
plane.  The motion planning of a convex polygonal robot in a polygonal
environment has been extensively studied (see for example
\cite{ks-empac-90,kst-coksm-97}).
Such algorithms can be
generalized  to plan the motion of a half-disk. It should lead to an
algorithm of complexity $O(n\lambda_s(n)\log n)$, where $\lambda_s(n)$
is an almost linear function of $n$.  The projection can be done using
classical techniques, such as projecting all the faces of \L\ and
computing their union.
  Since the complexity of the 3D object \L\ is
not directly related to the complexity of its projection, this
approach does not provide a combinatorial bound on $\F$.
However, assuming $|\F|=O(\lambda_s(|\A|))$ (which will be proved
in this paper) the time complexity of the algorithm 
of Kedem et al. is  $O(n\lambda_s(n)\log n+\lambda_s(|\A|)\log^2 n)$.

In this paper, we present a direct computation of $\F$.  This approach
provides an upper bound on the size of $\F$, namely
$|\F|=O(\lambda_s(|\A|))$. It also provides an algorithm
for computing \F\ in 
$O(\lambda_s(|\A|)\log n)$ time.  As in \cite{ss-nempa-87} and contrary
to \cite{kst-coksm-97}, the algorithm proposed here is sensitive to
$|\A|$ which is usually less than quadratic.  Another advantage of our
direct computation is to avoid the explicit construction of the 3D
object \L\ which is useless for our application. Our algorithm
manipulates only two-dimensional arrangements or lower envelopes and
we provide a detailed description of the curves involved in the construction.

Let us now detail the computation of \F\ in the
case of point footholds.  We know that each arc of the boundary \df\
of \F\ is either a straight line segment belonging to a line joining
two footholds or an arc of a circle $C_i$ (see
Section~\ref{Notations_and_previous_results}). The circular arcs
$\df\cap C_i$ are computed first
(Sections~\ref{Computation_of_df_cap_A}, \ref{Properties_of_zi}
and~\ref{Construction_of_df_cap_cio}) and linked together with the
line segments in a second step (Sections~\ref{cotaodfifaf}
and~\ref{construction_of_F}).

 \begin{figure}[p] \begin{center}
  \unitlength 1cm \input{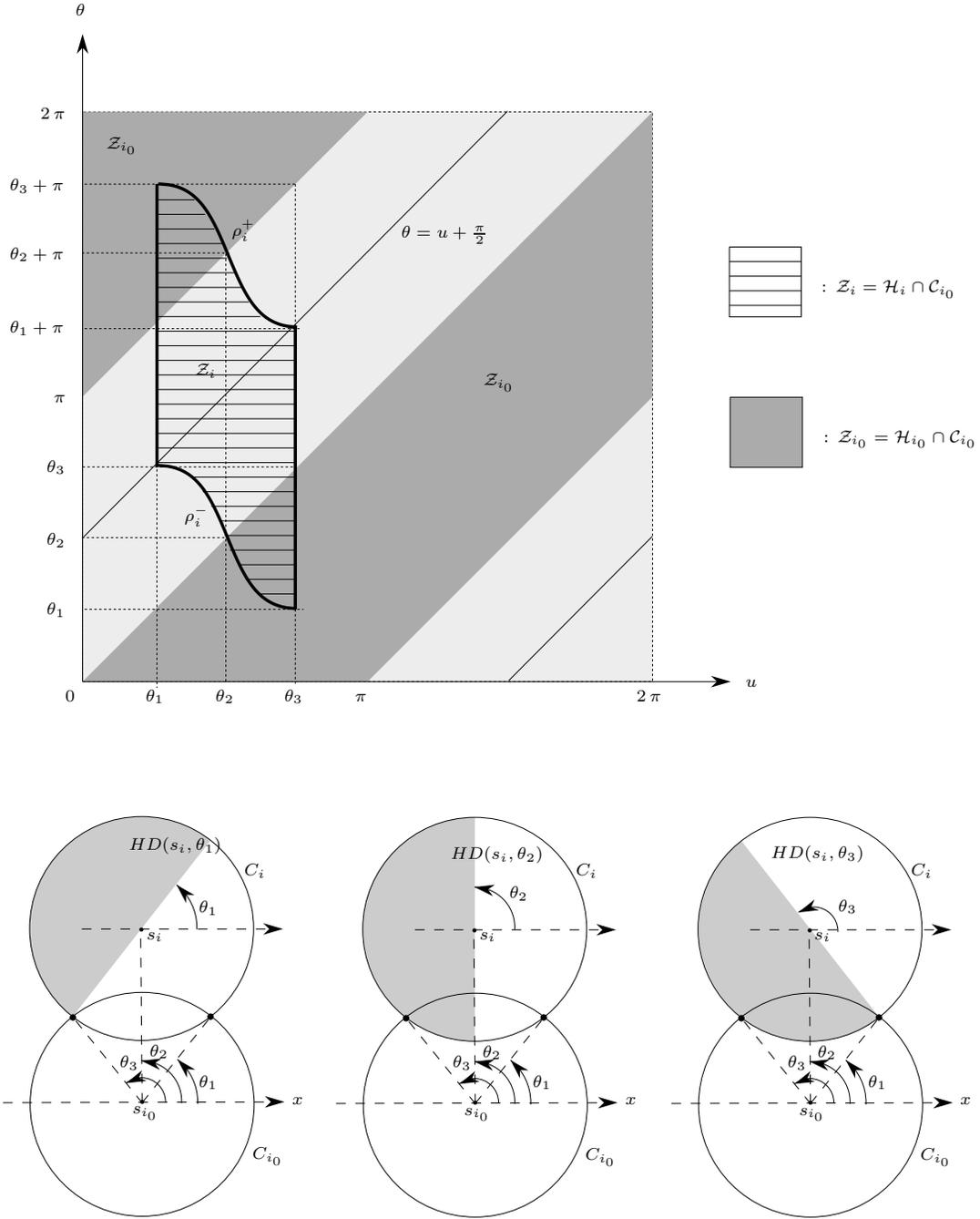} \end{center} \caption{Example of $\protect{\cal Z}_i$ for 
$\|s_{i_0}s_i\|=\protect\sqrt{2}\,R$ and some corresponding critical positions of~$HD(s_i,\theta)$.}
  \label{Example-Zi-and-corresponding-HD} \end{figure}

\subsection{Computation of $\df\cap\A$}
\label{Computation_of_df_cap_A}
 
In the sequel, the {\em contribution} of an object $X$ to another object $Y$ is $X\cap Y$. 
We compute the contribution of each circle $C_{i_0}$,
$i_0=1,\ldots,n$, to \df\ in turn.
Recall that \Co\ denote the torus $C_{i_0}\times S^1$.
The contribution of each circle $C_{i_0}$ to \df\ will be obtained by
computing the intersection of all the $\Hi$, $i=1,\ldots,n$, with the
torus $\Co$. Let $\Zi$, $i=1,\ldots,n$, denote these intersections:\\
\centerline{$\Zi = \Hi\cap\Co$.}

We first show how to compute the contribution of $C_{i_0}$ to \df\ in
term of the $\Zi$, and leave the studies of the shape and properties
of $\Zi$ to Section~\ref{Properties_of_zi}.  
Figures~\ref{Example-Zi-and-corresponding-HD} and
\ref{Construction-df-cap-Co-1-v2} show  some (hatched) 
$\Zi\subset\Co$ ($i \neq i_0$) where \Co\ is parameterized by
$(u,\theta)$ ($u$ and $\theta$ parameterize $C_{i_0}$ and $S^1$
respectively); the dark grey region shows ${\cal Z}_{i_0}$.
\begin{proposition}
\label{contribution}
The  contribution of $C_{i_0}$ to \df\ is:\\
\centerline{$C_{i_0} \cap \df =compl(p_{/\!/\theta}(compl(\cup_i\Zi)))\setminus int(compl(p_{/\!/\theta}(compl(\cup_{i\neq i_0}\Zi))))$.}
\end{proposition}
\proof 
Since \F\ is a closed set, $C_{i_0} \cap \df = [C_{i_0} \cap \F]
\setminus [C_{i_0} \cap int(\F)]$.  According to Theorem~\ref{thFHi},
$\F = compl(p_{/\!/\theta}(compl(\H)))$.  One can easily prove that
for any set $\E\in{\Re}^2\times S^1$,
$int(compl(\E))=compl(clos(\E))$, $clos(compl(\E))=compl(int(\E))$,
and $clos(p_{/\!/\theta}(\E))=p_{/\!/\theta}(clos(\E))$. It then
follows from the expression of $\F$ that
$int(\F)=compl(p_{/\!/\theta}(compl(int(\H))))$.

Recall that for any sets $X,Y\in {\Re}^2\times S^1$, $compl(X\cap Y)=compl(X)\cup compl(Y)$, 
$p_{/\!/\theta}(X\cup Y)=p_{/\!/\theta}(X)\cup p_{/\!/\theta}(Y)$, and 
$compl(X\cup Y)=compl(X)\cap compl(Y)$. That implies\\
\centerline{$compl(p_{/\!/\theta}(compl(X\cap Y)))=compl(p_{/\!/\theta}(compl(X)))\cap 
compl(p_{/\!/\theta}(compl(Y)))$.}
We now consider that equation with $X$ equal to $\H$ or $int(\H)$, and $Y$ equal to the torus $\Co$. 
Since $compl(p_{/\!/\theta}(compl(\Co)))$ is the circle $C_{i_0}$ we get: \\
\centerline{$compl(p_{/\!/\theta}(compl(\H\cap \Co)))=\F \cap C_{i_0}$ and}
\centerline{$compl(p_{/\!/\theta}(compl(int(\H)\cap \Co)))=int(\F) \cap C_{i_0}$.}

Since $\H=\cup_{i =1}^n\Hi$ and $\Zi = \Hi\cap\Co$ by definition,
$\H\cap\Co =\cup_{i =1}^n\Zi$ and $int(\H)\cap \Co = \cup_{i
=1}^n (int(\Hi)\cap\Co)$. By the general position assumption, no two
footholds lie at distance $2R$, thus for $i\neq i_0$,
$int(\H_{i})\cap{\cal C}_{i_0}=int(\Zi)$\footnote{Recall that $int$
denotes the relative interior, thus $int(\H_{i})$ is the interior of
\Hi\ in $\Re^2\times S^1$ but $int(\Z_{i})$ denotes the interior of
\Zi\ in $\Co$.}. As $int(\H_{i_0})\cap{\cal C}_{i_0}=\emptyset$, we get
$int(\H)\cap \Co = \cup_{i\neq i_0}int(\Zi)$.  The study of the
shape of  \Zi\ will yield (see Lemma~\ref{inter-int-Zi}) that
$\cup_{i\neq i_0}int(\Zi)=int(\cup_{i\neq i_0}\Zi)$. Therefore,
$int(\F)\cap C_{i_0}=compl(p_{/\!/\theta}(compl(int(\cup_{i\neq
i_0}\Zi))))=int(compl(p_{/\!/\theta}(compl(\cup_{i\neq i_0}\Zi))))$
and $\F\cap C_{i_0}=compl(p_{/\!/\theta}(compl(\cup_{i}\Zi)))$.  Using
$C_{i_0}
\cap \df = [C_{i_0} \cap \F]\setminus [C_{i_0} \cap int(\F)]$, we get
the result.
\endproof 

Thus, the contribution of $C_{i_0}$ to \df\ comes from the computation
of $\cup_{i}{\Zi}$ and $\cup_{i\neq i_0}{\Zi}$.

Geometrically, $compl(p_{/\!/\theta}(compl(\cup_i\Zi)))$ is the
vertical projection (along the $\theta$-axis) of the
largest vertical strip $\Sigma_{i_0}$ included in $\cup_{i}{\Zi}$ (see
Figure~\ref{Construction-df-cap-Co-1-v2}). Similarly,
$compl(p_{/\!/\theta}(compl(\cup_{i\neq i_0}\Zi)))$ is the projection
of the largest vertical strip $\Sigma'_{i_0}$ included in $\cup_{i\neq
i_0}\Zi$. Thus, $\df\cap C_{i_0}$ is the vertical projection onto
$C_{i_0}$ of the vertical strip $\Sigma_{i_0}\setminus int(\Sigma
'_{i_0})$.

In order to compute \F\ efficiently, we need to compute the union of
the regions $\Zi$ efficiently. More precisely, we will show that the
union of the regions $\Zi$ can be computed in $O(k_{i_0}\log k_{i_0})$
time where $k_{i_0}$ is the number of helicoidal volumes \Hi\
intersecting $\Co$.

This is possible because the \Zi\ have 
special shapes that allow us 
 to reduce the computation of their union to the computation of a
small number of lower envelopes of curves drawn on $\Co$, with the
property that two of them intersect at most once. The geometric
properties of the \Zi\ are discussed in Section \ref{Properties_of_zi}
and, in Section~\ref{Construction_of_df_cap_cio}, we present and
analyze the algorithm for constructing $\df\cap C_{i_0}$.

\subsection{Properties of the \Zi}
\label{Properties_of_zi}

We study here the regions $\Zi =\Hi\cap\Co$.  Recall that we
parameterize $\Co = C_{i_0}\times S^1$ by $(u,\theta)$ where $u$ and
$\theta$ parameterize $C_{i_0}$ and $S^1$ respectively ($u =0$
corresponds to the point of $C_{i_0}$ with maximum $x$-coordinate).
Figures~\ref{Example-Zi-and-corresponding-HD} and
\ref{Construction-df-cap-Co-1-v2} show examples of such regions
$\Zi$.
For convenience, we will use the vocabulary of the plane when
describing objects on the torus $\Co$. For instance, the curve drawn
on the torus \Co\ with equation 
$a \, \theta + b\, u + c =0$
will be called a line. The line $u=u_0$ will be called vertical and
oriented according to increasing $\theta$. Lower and upper will refer
to this orientation. The discussion below considers only non empty
regions $\Zi$ (such that $\|\so\si\|< 2R$).

 \begin{figure}[t] \begin{center}
  \unitlength 1cm \input{Construction-df-cap-Co-1-v2.ltex} \end{center} \caption{Contribution of $C_{i_0}$ to
$\df$ $(0<\|s_1\so\|<R$, $R\leq \|s_2\so\|<\protect\sqrt{2}\,R$,
$\protect\sqrt{2}\,R\leq \|s_3\so\|<2R)$.}
  \label{Construction-df-cap-Co-1-v2} \end{figure}

We introduce first some notations.  Let $HC_i(\theta)$ be the
half-circle of the boundary of $HD(\si ,\theta)$, \ie, 
$HC_i(\theta)=C_i\cap HD(\si ,\theta)$.  Let $r_i(\theta)$ be the
spoke of $C_i$ that makes an angle $\theta$ with the $x$-axis, \ie, 
$r_i(\theta)=\{s_i+\lambda \vec{u}_{\theta}\,|\ \lambda\in[0,R]\}$
where $\vec{u}_{\theta}$ is the unit vector whose polar angle is
$\theta$.  The boundary of \Hi\ is composed of the three following
patches:
\begin{eqnarray*}
\tti & = & \{ (HC_i(\theta), \theta) \in {\Re}^2\times S^1\} \\
\rrip & = & \{ (r_i (\theta ), \theta) \in {\Re}^2\times S^1\} \\
\rri ^- & = & \{ (r_i(\theta +\pi), \theta) \in {\Re}^2\times S^1\} 
\end{eqnarray*}

Let $\rho_i^-$ and $\rho_i^+$ denote the curves 
$\rri ^-\cap\Co$ and $\rri ^+\cap\Co$, respectively.  
Since $\rri^-$ and $\rri^+$ are  translated copies of one another, i.e., 
$\rri ^- = \rri ^+ \pm (0,0,\pi)$, we have: 
\begin{lemma}
\label{translated-copies}
$\rho_i^-$ and $\rho_i^+$ are translated copies of one another, i.e.,\\
$\rho_i^+ = \{(u,\theta)\in S^1\times S^1\,|\ (u,\theta-\pi)\in\rho_i^-\} = \{(u,\theta)\in S^1\times S^1\,|\ (u,\theta+\pi)\in \rho_i^-\}$.
\end{lemma} 
\begin{lemma}
\label{u-monotone}
The curves  $\rho_i^\pm$  are monotone in $u$.
\end{lemma} 
\proof
Assume for a contradiction that a curve $\rho_i^\pm$ is not monotone
in $u$. Then, there exists $u$ and $\theta \neq \theta'$ in $S^1$ such
that $(u,\theta)$ and $(u,\theta')$ parameterize points of
$\rho_i^\pm$.  By the definition of $\rri ^\pm$, it then follows that
the point $U\in C_{i_0}$ parameterized by $u$ belongs to the two spokes
$r_i(\theta)$ (or $r_i(\theta+\pi)$) and $r_i(\theta')$ (or
$r_i(\theta'+\pi)$).  
The intersection between any two of these spokes is exactly $\si$. 
Thus, $U=\si$, which contradicts (since $U\in C_{i_0}$) the general
position assumption saying that the distance between \si\ and \so\ is not
$R$.
\endproof

\begin{lemma}
\label{Zi0} 
The region $\Z_{i_0}$ is the subset of \Co\ 
parameterized by
$\{(u,\theta)\in S^1\times S^1\,|\ \theta\leq u\leq \theta+\pi\}$ 
(shown in  grey in Figures~\ref{Example-Zi-and-corresponding-HD}
 and \ref{Construction-df-cap-Co-1-v2}).
\end{lemma}
\proof  
For any $\theta\in S^1$, the intersection between $\H_{i_0}$ and the
``horizontal plane'' $\Pi_\theta$ is the half-disk
$HD(s_{i_0},\theta)$.
Similarly, the intersection between \Co\ and that plane is 
$C_{i_0}$. Thus, the intersection between $\Z_{i_0}$ and $\Pi_\theta$ is
$HC_{i_0}(\theta)$, which is parameterized on $C_{i_0}$ by $\{u\in
S^1\,|\ \theta\leq u\leq \theta+\pi\}$. That intersection is actually
on the plane $\Pi_\theta$ and is therefore parameterized on \Co\ by
$\{(u,\theta)\in S^1\times S^1\,|\ \theta\leq u\leq
\theta+\pi\}$.
\endproof

\begin{proposition}
\label{Zi} 
\Zi\ is a connected region bounded from below by 
$\rho_i^-$ and from above by $\rho_i^+$, i.e., 
$\Zi = \{(u,\theta)\in S^1\times S^1\,|\ \exists x\in[0,\pi], (u,\theta-x)\in \rho_i^-,  
(u,\theta-x+\pi)\in \rho_i^+\}$ (see Figures~\ref{Example-Zi-and-corresponding-HD}, \ref{Construction-df-cap-Co-1-v2}). 
\end{proposition}
\proof 
By cutting \Co\ and $\Hi$ by the ``horizontal plane'' $\Pi_\theta$, we
get that a point parameterized by $(u,\theta)$ on \Co\ belongs to
$\Hi$ if and only if the point $U$ parameterized by $u$ on $C_{i_0}$
belongs to $HD(s_i,\theta)$.  Since $HD(s_i,\theta)$ can be seen as
the union of the spokes $\{r_i(\theta+\gamma)\,|\ \gamma\in[0,\pi]\}$,
$(u,\theta)\in\Zi$ if and only if there exists $\gamma\in[0,\pi]$ such
that $U\in r_i(\theta+\gamma)$, or equivalently, $U\in r_i(\theta-x+\pi)$ with
$x=\pi-\gamma\in[0,\pi]$.  Since $\rri ^-
=\{(r_i(\theta-x+\pi),\theta-x)\,|\ \theta-x\in S^1\}$, it follows
from $U\in r_i(\theta-x+\pi)$ that the point of \Co\ parameterized by
$(u,\theta-x)$ belongs to $\rri ^-$ and thus to $\rho_i^-=\rri
^-\cap\Co$. From Lemma~\ref{translated-copies}, we get that the
point parameterized by $(u,\theta-x+\pi)$ belongs to $\rho_i ^+$.
Therefore, \Zi\ is a connected region bounded from below by $\rho_i^-$
and from above by $\rho_i^+$.
\endproof

We want to compute the union of the \Zi\ by computing the ``lower
envelope''\footnote{Note that the lower and upper envelopes of curves
in $S^1\times S^1$ are not actually defined.} of the lower edges
$\rho_i ^-$, and the ``upper envelope'' of the upper edges $\rho_i
^+$. It is unfortunately impossible to do so because some upper edges
$\rho_i ^+$ may possibly be ``below'' or intersect some lower edges
$\rho_j ^-$.  However, we can subdivide the regions \Zi\ into blocks
$\Zi ^k$, $k\in{\cal K}$, and separate these blocks into two sets
$\Omega_1$ and $\Omega_2$ such that the union of the $\Zi ^k$ in
$\Omega_1$ (resp. $\Omega_2$) is the region bounded from above by the
upper envelope of the upper edges of the $\Zi ^k\in\Omega_1$ and
bounded from below by the lower envelope of the lower edges of the
$\Zi ^k\in\Omega_1$ (resp. $\Omega_2$).  Such property can be realized
by showing that all the upper edges of the $\Zi ^k\in\Omega_1$ belong
to the strip $\{(u,\theta)\in
S^1\times[u+\frac{\pi}{2},u+\frac{3\pi}{2}]\}$ and all the lower edges
of the $\Zi ^k\in\Omega_1$ belong to the strip $\{(u,\theta)\in
S^1\times[u-\frac{\pi}{2},u+\frac{\pi}{2}]\}$ (a similar property is
shown for $\Omega_2$).  Note that the upper and lower envelopes are
then defined since they are considered in $S^1\times {\Re}$.

 \begin{figure}[th] \begin{center}
  \unitlength 1cm \input{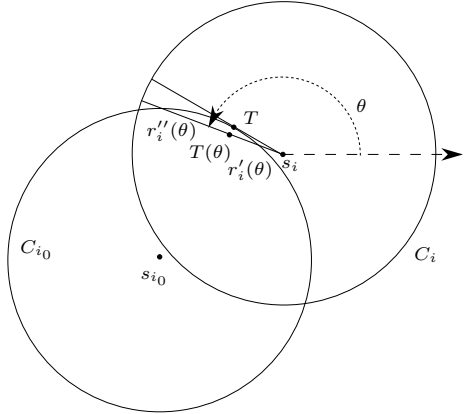} \end{center} \caption{For the definition of  $r_i'(\theta)$ and $r_i''(\theta)$.}
  \label{def-T-v2} \end{figure}

We subdivide \Zi\ into blocks $\Zi ^k$  when $R< \|\so\si\|<
\sqrt{2}\,R$. That subdivision is
performed such that the upper and lower edges of the $\Zi ^k$ are
$\theta$-monotone. Recall that the upper edge $\rho_i ^+$ of $\Zi$ is
the intersection of $\rri ^+=\{(r_i(\theta),\theta)\,|\ \theta\in
S^1\}$ and $\Co$.  The spoke $r_i(\theta)$ intersects $C_{i_0}$ twice
(for some $\theta$) when $R< \|\so\si\|< \sqrt{2}\,R$, which
implies that $\rho_i ^+$ is not $\theta$-monotone.  We cut the spoke
$r_i(\theta)$ into two pieces such that each piece intersects
$C_{i_0}$ at most once.  Let $T$ be the intersection point between
$C_{i_0}$ and on one of the two lines passing through \si\ and tangent
to $C_{i_0}$ (see Figure~\ref{def-T-v2}). Let $T(\theta)$ be the point
on $r_i(\theta)$ at distance $\|\si T\|$ from $\si$.  Cutting
$r_i(\theta)$ at $T(\theta )$ defines two sub-spokes $r_i'(\theta)$ and
$r_i''(\theta)$ that intersect $C_{i_0}$ in at most one point each;
without loss of generality, let $r_i'(\theta)$ denote the sub-spoke
joining $\si$ to $T(\theta )$.  The set of $\theta\in S^1$ for which
$r_i'(\theta)$ intersects $C_{i_0}$ is clearly connected but the set
of $\theta\in S^1$ for which $r_i''(\theta)$ intersects $C_{i_0}$
consists of two connected components.  We denote by $\rho_i ^{2+}$ the
intersection $\{(r_i'(\theta),\theta)\,|\ \theta\in S^1\}\cap\Co$ and
by $\rho_i ^{1+}$ and $\rho_i ^{3+}$ the two connected components of
the intersection $\{(r_i''(\theta),\theta)\,|\ \theta\in S^1\}\cap\Co$
(see Figure~\ref{Construction-df-cap-Co-1-v2}).  Since $r_i'(\theta)$
and $r_i''(\theta)$ intersect $C_{i_0}$ at most once for any
$\theta\in S^1$, the curves $\rho_i ^{1+}$, $\rho_i ^{2+}$ and $\rho_i
^{3+}$ are $\theta$-monotone.  The lower edges $\rho_i ^{k-}$,
$k=1,2,3$ can be defined similarly or in a simpler way as the
translated copies of $\rho_i ^{k+}$, $k=1,2,3$, i.e., $\rho_i^{k-} =
\{(u,\theta)\in S^1\times S^1\,|\ (u,\theta+\pi)\in\rho_i^{+k}\}$. We
denote by $\Zi ^k$, $k=1,2,3$, the subset of \Zi\ bounded from above
by $\rho_i ^{k+}$ and from below by $\rho_i ^{k-}$.

We can now prove the following proposition that will allow us to
compute the union of the \Zi\ by computing the upper and lower
envelopes of their upper and lower edges.

\begin{proposition}
\label{droitecoupe}
If $0 \leq \|\so\si\|< R$, the line $\theta =u-\frac{\pi}{2}$ properly
intersects $\Zi$, and the lines $\theta
=u\pm\frac{\pi}{2}$ properly intersect neither 
$\rho_i^{+}$ nor $\rho_i ^{-}$.

If $R< \|\so\si\|< \sqrt{2}\,R$, the line $\theta =u+\frac{\pi}{2}$
properly intersects $\Zi ^2$, and the line $\theta =u-\frac{\pi}{2}$
properly intersects $\Zi ^1$ and $\Zi ^3$.  Furthermore, the lines
$\theta =u\pm\frac{\pi}{2}$ properly intersect none of the edges
$\rho_i ^{1+}$, $\rho_i^{1-}$, $\rho_i ^{2+}$, $\rho_i^{2-}$, $\rho_i
^{3+}$ and $\rho_i^{3-}$.
 
If $\sqrt{2}\,R\leq \|\so\si\|<2R$, the line $\theta =u+\frac{\pi}{2}$
properly intersects $\Zi$, and the lines $\theta
=u\pm\frac{\pi}{2}$ properly intersect neither $\rho_i ^+$ nor $\rho_i
^-$.
\end{proposition}
\proof  
Let $(u_P,\theta_P)$ parameterize a point of a curve $\rho_i$. Let $P$
denote the point of $C_{i_0}$ with parameter $u_P$ and $\gamma =\angle
{(\overrightarrow{P\so},\overrightarrow{P\si})}\ [2\pi]$ (see
Figure~\ref{gamma-v2}).  One can easily show that $\gamma =\theta_P -u_P
[\pi]$.  We prove that $\gamma\neq \frac{\pi}{2}\ [\pi]$, except
possibly when $(u_P,\theta_P)$ is an endpoint of $\rho_i$ (or
$\rho_i^{k}$ when $R< \|\so\si\|< \sqrt{2}\,R$), which implies, since
$\gamma =\theta_P -u_P [\pi]$, that the lines $\theta
=u\pm\frac{\pi}{2}$ intersect neither $\rho_i^{+}$ nor $\rho_i^{-}$
(resp. $\rho_i^{k+}$ nor $\rho_i^{k-}$), except possibly at their
endpoints.
 \begin{figure}[th] \begin{center}
  \unitlength 1cm \input{gamma-v2.ltex} \end{center} \caption{For the proof of Proposition~\ref{droitecoupe}.}
  \label{gamma-v2} \end{figure}

\begin{figure}[th] \begin{center}
  \unitlength 1cm \input{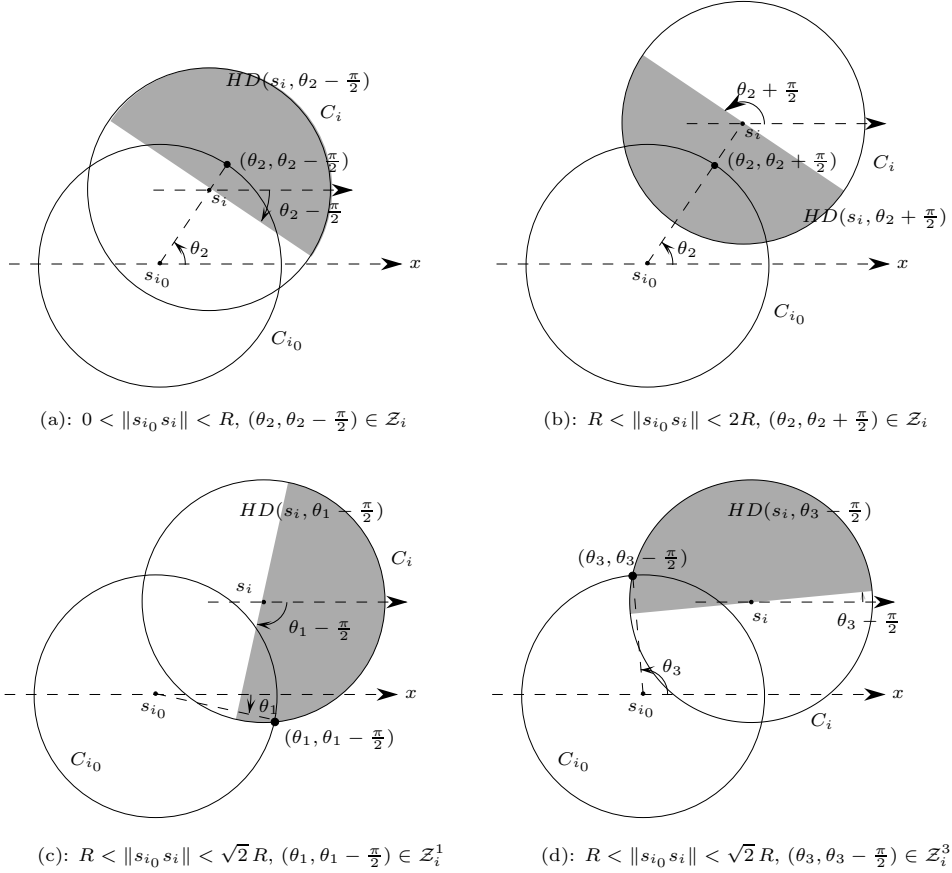} \end{center} \caption{For the proof of 
Proposition~\ref{droitecoupe}: section of $\Hi$ and $\Co$ by the
``planes'' $\Pi_{\theta_2-\frac{\pi}{2}}$,
$\Pi_{\theta_2+\frac{\pi}{2}}$, $\Pi_{\theta_1-\frac{\pi}{2}}$ and
$\Pi_{\theta_3-\frac{\pi}{2}}$ respectively.}
  \label{Points-dans-Zi} \end{figure}

{\bf Case 1}: $0 \leq \|\so\si\|< R$.  Since \si\ belongs to the disk of
radius $R$ centered at $\so$,
$\gamma\in(-\frac{\pi}{2},\frac{\pi}{2})$ for any $P\in C_{i_0}$ (see
Figure~\ref{gamma-v2}).  Thus, the lines $\theta =u\pm\frac{\pi}{2}$
properly intersect neither $\rho_i ^+$ nor $\rho_i ^-$.  Finally, the
point of \Co\ $(\theta_2,\theta_2-\frac{\pi}{2})$, where $\theta_2 =
\angle {(\vec{x},\overrightarrow{\so \si})}\ [2\pi]$, belongs to the
line $\theta =u-\frac{\pi}{2}$ and also to the relative interior of
\Zi\ since it belongs to the interior of \Hi\ (see
Figure~\ref{Points-dans-Zi}a).  Therefore, the line $\theta
=u-\frac{\pi}{2}$ properly intersects $\Zi$.

{\bf Case 2}: $R< \|\so\si\|< \sqrt{2}\,R$.  Let
$(u_{P_1},\theta_{P_1})$ parameterize the point connecting $\rho_i
^{1+}$ and $\rho_i ^{2+}$, and $(u_{P_2},\theta_{P_2})$ parameterize
the point connecting $\rho_i ^{2+}$ and $\rho_i ^{3+}$.  Let $P_1$ and
$P_2$ denote the points of $C_{i_0}$ parameterized by $u_{P_1}$ and
$u_{P_2}$ respectively.  According to the construction of $\rho_i
^{1+}$, $\rho_i ^{2+}$ and $\rho_i ^{3+}$, the tangent lines to
$C_{i_0}$ at $P_1$ and $P_2$ pass through $s_i$.  At most two tangent
lines to $C_{i_0}$ pass through $s_i$, thus $P_1$ and $P_2$ are the
only points of $C_{i_0}$ where $\gamma =
\frac{\pi}{2}\ [\pi]$.  
Since $\rho_i ^+$ is $u$-monotone by Lemma~\ref{u-monotone}, $(u_{P_1},\theta_{P_1})$ 
and $(u_{P_2},\theta_{P_2})$ are the only points of $\rho_i ^+$ where $\gamma =
\frac{\pi}{2}\ [\pi]$.  
Therefore, the lines $\theta= u \pm \frac {\pi}{2}$ do not properly
intersect $\rho_i ^{k+}$, $k=1,2,3$.  Similarly, the lines $\theta= u
\pm \frac {\pi}{2}$ do not properly intersect $\rho_i ^{k-}$,
$k=1,2,3$. 

Let $\theta_1$ and $\theta_3$ be the parameters on $C_{i_0}$ of the
intersection points between $C_{i_0}$ and $C_i$ (see
Figures~\ref{Points-dans-Zi}{c} and d); to differentiate $\theta_1$
from $\theta_3$, assume without loss of generality that, for any
$\varepsilon>0$ small enough, the points of $C_{i_0}$ parameterized by
$\theta_1+\varepsilon$ and $\theta_3-\varepsilon$ are in the disk of
radius $R$ centered at $\si$.  Then, the points
$(\theta_1,\theta_1-\frac{\pi}{2})$ and
$(\theta_3,\theta_3-\frac{\pi}{2})$ of $\Co$ belong to $\Zi ^1$ and
$\Zi ^3$ (or to $\Zi ^3$ and $\Zi ^1$) respectively (see
Figures~\ref{Points-dans-Zi}{c} and {d}).
However, these points do not belong to the relative interior of $\Zi
^1$ and $\Zi ^3$ (because they lie on the border of
$HD(\si,\theta_1-\frac{\pi}{2})$ and
$HD(\si,\theta_3-\frac{\pi}{2})$). Nevertheless, there clearly exists
$\varepsilon>0$ small enough such that the point parameterized by
$\theta_1+\varepsilon$ (resp. $\theta_3-\varepsilon$) on $C_{i_0}$
belongs to the interior of the half-disk $HD(\si,
\theta_1-\frac{\pi}{2}+\varepsilon)$ (resp. $HD(\si,
\theta_3-\frac{\pi}{2}-\varepsilon)$).  Thus, the points
$(\theta_1+\varepsilon,\theta_1+\varepsilon-\frac{\pi}{2})$ and
$(\theta_3-\varepsilon,\theta_3-\varepsilon-\frac{\pi}{2})$ of $\Co$
belong to the relative interior of $\Zi ^1$ and $\Zi ^3$ respectively.
Therefore, the line $\theta =u-\frac{\pi}{2}$ properly intersects $\Zi
^1$ and $\Zi ^3$.

On the other hand, $(\theta_2,\theta_2+\frac{\pi}{2})$ (where
$\theta_2 = \angle {(\vec{x},\overrightarrow{\so \si})}\ [2\pi]$)
belongs to relative interior of $\Zi ^2$ because the point of
$C_{i_0}$ parameterized by $\theta_2$ belongs to the relative interior
of the sub-spoke $r_i'(\theta_2+\pi)$ (see
Figure~\ref{Points-dans-Zi}b) which belongs to interior of
$HD(\si,\theta_2+\frac{\pi}{2})$.  Therefore, the line $\theta
=u+\frac{\pi}{2}$ properly intersects $\Zi ^2$.

{\bf Case 3}: $\sqrt{2}\,R\leq \|\so\si\|<2R$. Since $r_i(\theta)$
intersects $C_{i_0}$ at most once, $\gamma\in [\frac{\pi}{2},
\frac{3\pi}{2}]$ (see Figure~\ref{gamma-v2}). Moreover,
$\gamma=\frac{\pi}{2}$ $[\pi ]$ only when $\|\so\si\|=\sqrt{2}\,R$,
but then, $P$ is at distance $R$ from \si\ which implies that
$(u_P,\theta_P)$ is an endpoint of $\rho_i$.  Thus, the lines
$\theta=u\pm\frac{\pi}{2}$ intersect neither $\rho_i ^+$ nor $\rho_i
^-$, except possibly at their endpoints.  Finally, the point 
$(\theta_2,\theta_2+\frac{\pi}{2})$ of \Co\ (where $\theta_2 = \angle
{(\vec{x},\overrightarrow{\so \si})}\ [2\pi]$) belongs to the line
$\theta =u+\frac{\pi}{2}$ and also to the relative interior of \Zi\
(see Figures~\ref{Points-dans-Zi}b and
\ref{Example-Zi-and-corresponding-HD}). Therefore, the line $\theta
=u+\frac{\pi}{2}$ properly intersects $\Zi$.
\endproof 

By Proposition~\ref{droitecoupe}, we can compute the union
$\cup_{i\neq i_0}\Zi$ by separating the $\Zi$, $\Zi ^k$ into two sets
$\Omega_1$ and $\Omega_2$ (where $\Zi$, $\Zi ^k$ belongs to $\Omega_1$
if and only if $\rho_i ^+$, $\rho_i ^{k+}$ belongs to the strip
$\{(u,\theta)\in S^1\times[u+\frac{\pi}{2},u+\frac{3\pi}{2}]\}$ and
$\rho_i ^-$, $\rho_i ^{k-}$ belongs to the strip $\{(u,\theta)\in
S^1\times[u-\frac{\pi}{2},u+\frac{\pi}{2}]\}$) and computing the union
of the $\Zi$, $\Zi ^k$ in $\Omega_1$ (resp. $\Omega_2$) by computing
the upper envelope of their upper edges and the lower envelope of
their lower edges. In order to compute efficiently these upper and
lower envelopes, we show that the curves $\rho_i ^+$, $\rho_i ^-$,
$\rho_i ^{k+}$ and $\rho_i ^{k-}$ intersect each other at most once.
However, we need for that purpose to split the regions \Zi\ when $0 <
\|\so\si\|< R$ into two blocks $\Zi ^1$ and $\Zi ^2$ separated by 
the vertical line $u =\theta_2= \angle {(\vec{x},\overrightarrow{\so
\si})}$; it also remains to split the $\theta$-interval (or the
$u$-interval) over which $\rho_i$ is defined into two intervals of
equal length over which $\rho_i ^{1\pm}$ and $\rho_i ^{2\pm}$ are
defined (see Figure~\ref{Construction-df-cap-Co-1-v2}). Note that
Proposition~\ref{droitecoupe} still holds if we replace (when $0
<\|\so\si\|< R$) \Zi\ by $\Zi ^k$ and $\rho_i ^{\pm}$ by $\rho_i
^{k\pm}$, $k=1,2$.

For consistency, we split $\Z_{i_0}$ into two blocks $\Z_{i_0}^1$ and
$\Z_{i_0}^2$ separated by a vertical line (chosen arbitrarily, say $u
=\pi$). Also for consistency, the curves $\rho_i ^\pm$ when
$\sqrt{2}\,R\leq \|\so\si\|<2R$ are occasionally denoted in the sequel
$\rho_i ^{1\pm}$.

\begin{lemma}
\label{unefois} 
Let $\rho_i'$ and $\rho_j'$ be some connected portions of $\rho_i
^\pm$ and $\rho_j ^\pm$ respectively ($i\neq j$).  If $\rho_i'$ or
$\rho_j'$ is monotone in $\theta$ and defined over a $\theta$-interval
smaller than $\pi$, then $\rho_i'$ and $\rho_j'$ intersect at most
once.
\end{lemma}
\proof  
Let $(u_I,\theta_I)$ be a point of intersection between $\rho_i'$ and
$\rho_j'$ and $I$ be the point of the circle \co\ with parameter
$u_I$.  Since $\rho_i'$ is a portion of the intersection between \Co\
and $\rri ^\pm$, ${I}$ is a point of intersection between $C_{i_0}$ and the
diameter of $HD(\si,\theta_I)$.  Therefore, the line passing through
\si\ and $I$ has slope $\theta_I$.

By applying the same argument to $\rho_j'$, we obtain that \si\ and
\sj\ belong to the same straight line of slope $\theta_I$.  Therefore,
if $\rho_i'$ and $\rho_j'$ intersect twice, at $(u_I,\theta_I)$ and
$(u_J,\theta_J)$, then $\theta_I = \theta_J[\pi]$.  
It follows, if $\rho_i'$ or
$\rho_j'$ is defined over a $\theta$-interval smaller than $\pi$, 
that
$\theta_I= \theta_J[2\pi]$.  Furthermore, if $\rho_i'$ or $\rho_j'$ is
monotone in $\theta$, then $(u_I,\theta_I)$ and $(u_J,\theta_I)$ are
equal.
\endproof

\begin{lemma}
\label{inter-int-Zi} 
$\forall i,j$, $int(\Zi)\cup int(\Z_j)=int(\Zi\cup\Z_j)$.
\end{lemma}
\proof  
We assume that $i\neq j$ because otherwise the result is trivial. 
One can easily show that $int(\Zi)\cup int(\Z_j)\neq int(\Zi\cup\Z_j)$
only if the boundaries of \Zi\ and $\Z_j$ partially coincide, \ie, the
dimension of $\partial(\Zi)\cap\partial(\Z_j)$ is 1.  

By Proposition~\ref{Zi}, $\partial(\Zi)$ consists of the edges $\rho_i
^+$ and $\rho_i ^-$ and of two vertical line segments joining the
endpoints of $\rho_i ^+$ and $\rho_i ^-$ when these endpoints exist
(which is the case when $i\neq i_0$).
Moreover, these vertical line segments are clearly supported by the
vertical lines $u = \theta_1$ and $u = \theta_3$ where $\theta_1$ and
$\theta_3$ parameterize on $C_{i_0}$ the points of intersection
between $C_{i_0}$ and $C_i$ (see
Figure~\ref{Example-Zi-and-corresponding-HD}).

By Lemma~\ref{unefois}, the edges $\rho_i ^\pm$ and $\rho_j ^\pm$ do
not partially coincide.  By the general position assumption, 
no three distinct circles $C_{i_0}$, $C_i$ and $C_j$ have a common
intersection point.  Thus, for any $i\neq j$, $C_{i_0}\cap C_i$ and
$C_{i_0}\cap C_j$ are disjoint. Therefore, the vertical lines
$\partial(\Zi)\setminus\{\rho_i ^+,\rho_i ^-\}$ and
$\partial(\Z_j)\setminus\{\rho_j ^+,\rho_j ^-\}$ do not partially
coincide.  Finally, since $\rho_i ^\pm$ is nowhere partially supported
by a vertical line by Lemma~\ref{u-monotone}, $\rho_i ^\pm$ and the
vertical lines $\partial(\Z_j)\setminus\{\rho_j ^+,\rho_j ^-\}$ do not
partially coincide.
\endproof

\begin{proposition}
\label{une-intersection}
Any two curves among the curves $\rho_i ^{k\pm}$ intersect at most
once (where $k\in\{1,2\}$ if $0 \leq \|\so\si\|< R$, $k\in\{1,2,3\}$
if $R< \|\so\si\|<
\sqrt{2}\,R$, and $k=1$ if $\sqrt{2}\,R\leq \|\so\si\|<2R$).
\end{proposition}
\proof
By Lemma~\ref{unefois}, it is sufficient to prove that all the curves
$\rho_i ^{k\pm}$, $i\neq i_0$, are monotone in $\theta$ and defined
over $\theta$-intervals smaller than $\pi$. Indeed, the curves
$\rho_{i_0} ^{1+}$, $\rho_{i_0} ^{1-}$, $\rho_{i_0} ^{2+}$ and
$\rho_{i_0} ^{2-}$  clearly do not pairwise intersect more
than once, by Lemma~\ref{Zi0}.

If $0< \|\so\si\| < R$, any spoke of $C_i$ intersects $C_{i_0}$ at
most once. Hence, $\rho_i ^\pm$ is monotone in $\theta$. $\rho_i ^\pm$
is defined over a $\theta$-interval greater than $\pi$ but smaller
than $2\pi$. Since we have split that interval in two equal parts,
$\rho_i ^{1\pm}$ and $\rho_i ^{2\pm}$ are defined over a
$\theta$-interval smaller than $\pi$ (see $\Z_1$ in Figure
\ref{Construction-df-cap-Co-1-v2}).

If $R< \|\so\si\|< \sqrt{2}\,R$, the $\theta$-interval
where $r_i(\theta)$ (or $r_i (\theta+\pi)$) intersects $C_{i_0}$ is
smaller than $\pi$, which implies that $\rho_i$ is defined over a
$\theta$-interval smaller than $\pi$.  The curves $\rho_i ^{k+}$,
$k=1,2,3$, are defined as the connected components of
$\{(r_i'(\theta),\theta)\,|\ \theta\in S^1\}\cap\Co$ and
$\{(r_i''(\theta),\theta)\,|\ \theta\in S^1\}\cap\Co$. Since the
sub-spokes $r_i'(\theta)$ and $r_i''(\theta)$ intersect  $C_{i_0}$ at
most once for any $\theta\in S^1$, the curves $\rho_i ^{k+}$,
$k=1,2,3$, are $\theta$-monotone.

If $\sqrt{2}\,R\leq \|\so\si\|<2R$, $r_i(\theta)$ (and also $r_i
(\theta+\pi)$) intersects $C_{i_0}$ in at most one point, which proves
that $\rho_i$ is monotone in $\theta$.  Furthermore, the
$\theta$-interval where $\rho_i$ is defined is smaller than $\pi$
because the $\theta$-interval where $r_i(\theta)$ (or $r_i
(\theta+\pi)$) intersects $C_{i_0}$ is smaller than $\pi$.
\endproof

\subsection{Construction of $\df\cap C_{i_0}$}
\label{Construction_of_df_cap_cio}

We first show how to compute $\displaystyle\cup_{i}{\Zi}$. Let
$\Omega_1$ and $\Omega_2$ be the following sets of $\Zi ^k$:\\
\centerline{$\Omega_1=\{\Zi\,|\ \sqrt{2}\,R\leq \|\so\si\|<2R\} \cup 
\{\Zi ^2\,|\ R <\|\so\si\|< \sqrt{2}\,R \}$,}  \\
\centerline{$\Omega_2=\{\Zi ^1, \Zi ^2\,|\ 0 \leq \|\so\si\|< R\}\cup
\{\Zi ^1, \Zi^3\,|\ R < \|\so\si\|< \sqrt{2}\,R\}$.}

By Proposition~\ref{droitecoupe}, the line $\theta
=u+\frac{\pi}{2}$ properly intersects all the $\Zi ^k\in\Omega_1$ but
the lines $\theta=u\pm\frac{\pi}{2}$ properly intersect none of their
upper and lower edges $\rho_i ^{k+}$ and $\rho_i ^{k-}$.  Thus, the
regions $\Zi ^k\in\Omega_1$ can be seen as regions of $\{(u,\theta)\in
S^1\times[u-\frac{\pi}{2},u+\frac{3\pi}{2}]\}$ such that all their
upper edges $\rho_i ^{k+}$ lie in $\{(u,\theta)\in
S^1\times[u+\frac{\pi}{2},u+\frac{3\pi}{2}]\}$ and all their lower
edges $\rho_i ^{k-}$ lie in $\{(u,\theta)\in
S^1\times[u-\frac{\pi}{2},u+\frac{\pi}{2}]\}$.  Therefore, 
the union of the $\Zi ^k \in\Omega_1$ is the region of
$\{(u,\theta)\in S^1\times[u-\frac{\pi}{2},u+\frac{3\pi}{2}]\}$
bounded from above by the upper envelope of their $\rho_i ^{k+}$ and
bounded from below by the lower envelope of their $\rho_i ^{k-}$.
Similarly, the union of the $\Zi ^k \in\Omega_2$ is the
region of $\{(u,\theta)\in
S^1\times[u-\frac{3\pi}{2},u+\frac{\pi}{2}]\}$ bounded from above by
the upper envelope of the $\rho_i ^{k+}$ and bounded from below by the
lower envelope of the $\rho_i ^{k-}$.

The union of $\Omega_1$ and $\Omega_2$, which is $\displaystyle
\cup_{i}{\Zi}$, can be  achieved by computing, on one hand, 
the intersection between the upper edge chain of
$\displaystyle\cup_{\Zi ^k\in\Omega_1}{\Zi ^k}$ with the lower edge
chain of $\displaystyle\cup_{\Zi ^k\in\Omega_2}{\Zi ^k}$ (which both
belong to $\{(u,\theta)\in S^1\times S^1 \,|\
\theta\in[u+\frac{\pi}{2},u+\frac{3\pi}{2}]\}$), and on the other
hand, the intersection between the upper edge chain of
$\displaystyle\cup_{\Zi ^k\in\Omega_2}{\Zi ^k}$ with the lower edge
chain of $\displaystyle\cup_{\Zi ^k\in\Omega_1}{\Zi ^k}$ (which both
belong to $\{(u,\theta)\in S^1\times S^1 \,|\
\theta\in[u-\frac{\pi}{2},u+\frac{\pi}{2}]\}$).  These intersections
can simply be performed by following the two edge chains for $u$ from
$0$ to $2\pi$, since they are monotone in $u$ by Lemma~\ref{u-monotone}.

Let us analyze the complexity of the above construction. The \ko\
helicoidal volumes \Hi\ that intersect \Co\ can be found in $O(\ko)$
amortized time once the Delaunay triangulation of the footholds has
been computed, which can be done in $O(n\log n)$
time~\cite{dd-frspp-90,t-frnns-91}.  By
Proposition~\ref{une-intersection}, the upper and lower envelopes can
be computed in $O(k_{i_0}\log k_{i_0})$ time using
$O(k_{i_0}\alpha(k_{i_0}))$ space where $\alpha$ is the pseudo inverse
of the Ackerman's function~\cite{h-fuenl-89}.  Also by
Proposition~\ref{une-intersection}, the union of $\Om_1$ and $\Om_2$
can be done in linear time in the size of the edge chains, that is
$O(k_{i_0}\alpha(k_{i_0}))$ time. 
Thus, we can compute $\displaystyle
\cup_{i}{\Zi}$ in  $O(k_{i_0}\log k_{i_0})$ time using
$O(k_{i_0}\alpha(k_{i_0}))$ space after $O(n\log n)$ preprocessing
time. We can compute $\displaystyle\cup_{i\neq i_0}{\Zi}$ similarly by
removing $\Z_{i_0}^1$ and $\Z_{i_0}^2$ from $\Om_2$.

The contribution of $C_{i_0}$ to \df\ is, according to
Proposition~\ref{contribution}, $C_{i_0} \cap \df
=compl(p_{/\!/\theta}(compl(\cup_i\Zi)))\setminus
int(compl(p_{/\!/\theta}(compl(\cup_{i\neq i_0}\Zi))))$.  By
Remark~\ref{RemFHi}, 
\\ $compl(p_{/\!/\theta}(compl(\cup_i\Zi)))$ and
$compl(p_{/\!/\theta}(compl(\cup_{i\neq i_0}\Zi)))$ are the
projections onto $C_{i_0}$ of the largest vertical strips $\Sigma
_{i_0}$ and $\Sigma '_{i_0}$ included in $\displaystyle \cup_{i}\Zi$
and $\displaystyle \cup_{i\neq i_0}\Zi$, respectively (see
Figure~\ref{Construction-df-cap-Co-1-v2}). These projections are
easily computed because the edges of $\displaystyle \cup_{i}\Zi$ and
$\displaystyle \cup_{i\neq i_0}\Zi$ are monotone with respect to $u$
(Lemma~\ref{u-monotone}). These projections, and therefore the
computation of $C_{i_0} \cap \df$, can thus be done in linear time and
space in the size of $\displaystyle \cup_{i}\Zi$ and $\displaystyle
\cup_{i\neq i_0}\Zi$, that is $O(k_{i_0}\alpha(k_{i_0}))$.

Moreover, we label an arc of \df\ either by $i$ if the arc belongs to
the circle $C_i$ or by $(i,j)$ if the arc belongs to the straight line
segment $[s_i,s_j]$. The labels of the edges of \df\ incident to
$C_{i_0}$ can be found as follows, without increasing the
complexity. An arc of $\df\cap C_{i_0}$ corresponds to a vertical
strip $\Sigma_{i_0}\setminus \Sigma '_{i_0}$.  An endpoint $P$ of such an
arc is the projection of a vertical edge, or the
projection of a point of intersection between two curved edges.  In
the first case, $P$ is the intersection of \co\ with some \ci\ and in
the second case, $P$ is the intersection of \co\ with some line
segment $[\si,\sj]$.  By the general position assumption, among the
circles $C_1,\ldots,C_n$ and the line segments joining two footholds,
the intersection between three circles or, two circles and a line
segment or, one circle and two line segments, is empty.  Thus, $P$ is
the intersection between \co\ and either a unique \ci\ or a unique
line segment $[\si,\sj]$.  Therefore, the edge of \df\ incident to
\co\ at $P$ is either a circular arc supported by \ci\ or a line 
segment supported by $[\si,\sj]$.  Hence, the labels of the edges of
\df\ incident to $C_{i_0}$ can be found at no extra-cost during the
construction.

Since \A\ is the arrangement of the circles of radius $R$ centered at
the footholds, $\sum_{i_0 =1}^n k_{i_0}=O(|\A|)$. The above
considerations yield the following theorem:

\begin{theorem}
\label{R12} 
We can compute $\df\cap \A$ and the labels of the edges of
\df\ incident to the arcs of $\df\cap \A$ in $O(|\A|\log n)$ time using
$O(|\A|\alpha(n))$ space.
\end{theorem}

\subsection{Computation of the arcs of \df\ issued from a foothold}
\label{cotaodfifaf}

The previous section has shown how to compute all the vertices of \F\
that are incident to at least one circular arc. It remains to find the
vertices of \F\ incident to two straight edges.  As we have seen in
Section~\ref{Notations_and_previous_results},  a vertex of $\F$
incident to two straight edges of $\df$ is a foothold. Furthermore,
considering a foothold $s_{i_0}$ in a cell $\Gamma$ of $\A$, $s_{i_0}$
is a vertex of $\F$ incident to two straight edges of $\df$ if and only
if $s_{i_0}$ is a vertex of the convex hull of the footholds reachable
from $s_{i_0}$.  The $k_{i_0}'$ footholds contained in the disk
of radius $R$ centered at $\so$  can be found in
$O(k_{i_0}')$ amortized time because we have already computed the
Delaunay triangulation of the
footholds~\cite{dd-frspp-90,t-frnns-91}. Thus, we can decide if \so\
is a vertex of the convex hull of these $k_{i_0}'$ footholds in
$O(k_{i_0}')$ time and space. When \so\ is a vertex of the convex
hull, we can also find the two edges of the convex hull adjacent to
\so\ in $O(k_{i_0}')$ time and space.  As the sum of
the $k_i'$ for $i\in\n$ is bounded by the size of $\A$, we obtain the
following theorem:
\begin{theorem}
\label{R34} 
The footholds belonging to \df\ and the labels of the arcs of \df\
issued from these footholds can be found in $O(|\A|)$ time and space.
\end{theorem}

\subsection{Construction of \F}
\label{construction_of_F}

~

\begin{theorem}
\label{R} 
The free space of the spider robot can be computed in $O(|\A|\log n)$
time using $O(|\A|\alpha(n))$ space.
\end{theorem}
\proof  
By Theorem~\ref{R12}, we have computed all the circular arcs of \df\
and the labels of the edges of \df\ incident to them. By
Theorem~\ref{R34}, we have computed all the vertices of \df\ that are
incident to two straight edges of \df\ and the label of these two
edges.  It remains to sort the vertices of \df\ that appear on the
line segments $[\si,\sj]$.  We only consider the line segments
$[\si,\sj]$ such that the corresponding label $(i,j)$ appears during
previous computations.  Then, we sort the vertices of \df\ that belong
to each such relevant line.  Since
$|\df|=\Theta(|\A|)$~\cite{bddp-mplrs-95}, sorting all these vertices  can be
done in $O(|\A|\log n)$ time.  
A complete description of $\df$ then follows easily.
\endproof

\section{Generalization to polygonal foothold regions}
\label{polygonal-foothold-regions}

\subsection{Introduction and preliminaries}
\label{Introduction-and-preliminaries}

We consider now the case where the set of footholds is no longer a set
of points but a set \S\ of pairwise disjoint polygonal regions bounded
by $n$ line segments $e_1,\ldots,e_n$.  Clearly, \S\ is a subset of
the free space \F\ of the spider robot.  Let \Fp\ denote the free
space of the spider robot using as foothold regions only the edges
$e_1,\ldots,e_n$. Suppose that the spider robot admits a stable
placement outside \S\ with its feet inside some polygonal footholds;
then the placement remains stable if it retracts its legs on the
boundary of these polygonal regions. Hence, $\F=\F_e\cup\S$. We show
how to compute~$\Fp$.

As observed in Remark \ref{RemGeneralize},
 the results of Section~\ref{fsrthdr} remains true if the foothold
regions are line segments provided that \Hi\ is replaced by \Hei\ the
generalized helicoidal volume defined by (see Figure~\ref{EHi}):
\[\Hei = \{(P,\theta)\in\Re^2\times S^1\,|\ P\in HD(s,\theta),\ s\in\ei\}.\]
The helicoidal volume associated to a point site \si\ will be,
henceforth, denoted by $\Hsi$.
 \begin{figure}[ht] \begin{center}
  \unitlength .7cm \input{EHi.ltex} \end{center} \caption{Section of $\Hei$ by the ``plane'' $\Pi_{\theta}$.}
  \label{EHi} \end{figure}

Similarly, we define the generalized circle \cei\ as the set of points
at distance $R$ from $\ei$.  Let \Ae\ denote the arrangement of the $n$
generalized circles $C_{e_1},\ldots,C_{e_n}$.  Notice that
$|\Ae|=\Theta(n^2)$.

Each arc of the boundary \dfp\ of \Fp\ is either an arc of \cei\ 
corresponding to a maximal extension of one leg, or an arc
corresponding to placements at the limit of stability of the spider
robot.  Similarly to what we did in Section~\ref{Computation_of_F}, we
compute first the contribution of each \cei\ to \dfp\
(Sections~\ref{Computation_of_dfp_cap_EA}).  Thereafter, we compute
the arcs of \dfp\ that correspond to placements where the spider robot
is at the limit of stability (Section~\ref{unstable-equilibrium}).  
Finally, we show how to construct \Fp\ (and $\calf$) in
Section~\ref{Construction_of_F_2}.

\begin{figure}[tp] \begin{center}
  \unitlength 1cm \input{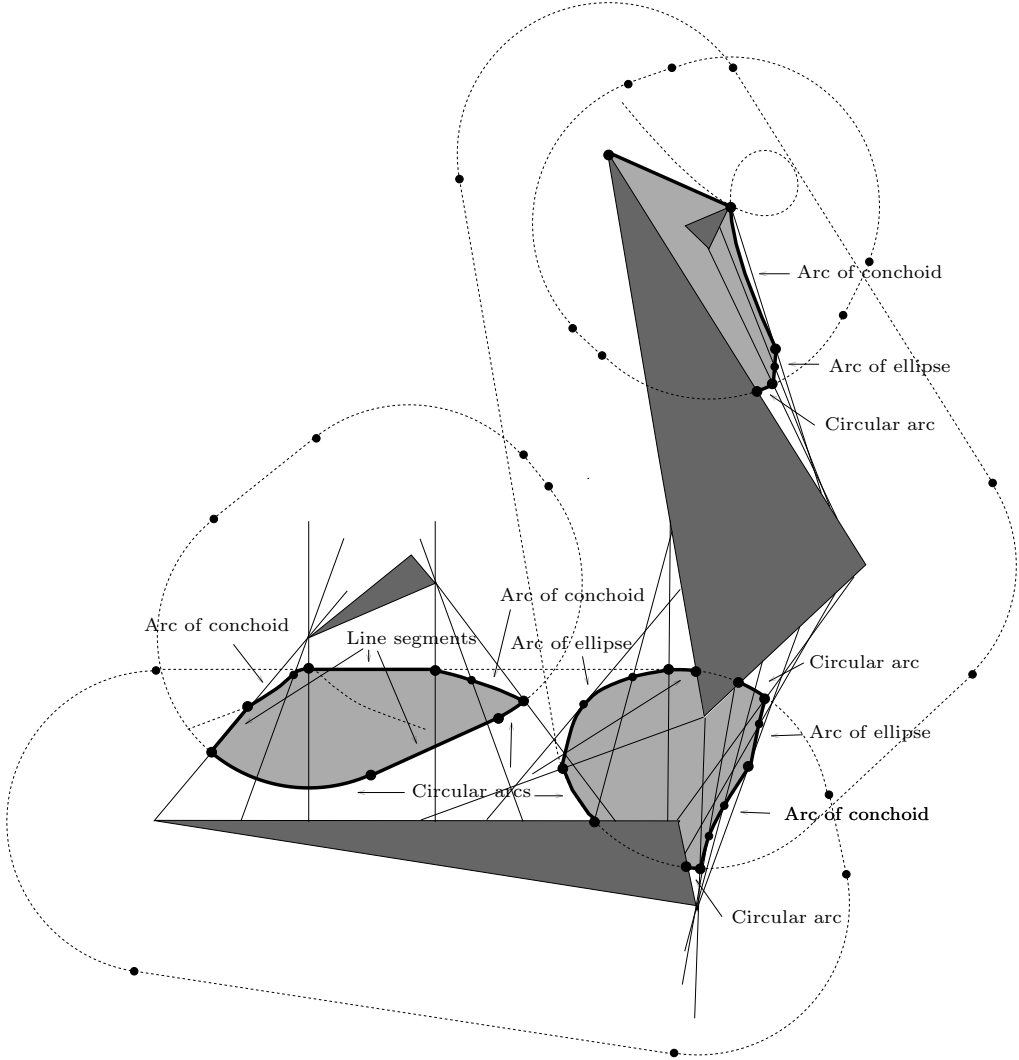} \end{center} \caption{Example of free space \Fp\ for polygonal foothold regions. 
The polygonal foothold regions are shown in dark grey. 
The other parts of \Fp\ are in light grey. The $\cei$ and some arcs of conchoid are dashed. 
All the line segments touching the polygons in 
two points are of length $2R$ and represent the ladder introduced in Section~\ref{unstable-equilibrium}.}
  \label{F-poly} \end{figure}

Figure~\ref{F-poly} shows an example of free space \Fp\ for polygonal foothold regions. 

\subsection{Computation of $\dfp\cap\Ae$}
\label{Computation_of_dfp_cap_EA}

We compute the contribution to \dfp\ of each generalized circle \cei\
in turn.  We consider the contribution of \ceio\ to \dfp\ for some
$i_0\in\{1,\ldots,n\}$.
\ceio\ is composed of two half-circles and two straight line
segments. In order to compute the contribution of \ceio\ to $\dfp$, we
evaluate first the contribution of the half-circles and then the
contribution of the straight line segments. For convenience, we will
not compute the contribution of the half-circles to \dfp\ but the
contribution of the whole circles.  Similarly, we will compute the
contribution of the whole straight lines supporting the line segments
of $\ceio$.

Let $s_{i_0}$ and $s'_{i_0}$ denote the two endpoints of the line
segment $e_{i_0}$, and let \cso\ and $C_{s'_{i_0}}$ denote the unit
circles centered at $s_{i_0}$ and $s'_{i_0}$ respectively.  Let
$l_{i_0}$ and $l'_{i_0}$ denote the two straight line segments of
$C_{e_{i_0}}$, and  $L_{i_0}$ and $L'_{i_0}$ their supporting
lines.  We show how to compute the contributions of \cso\ and
$L_{i_0}$ to $\dfp$; the contributions of $C_{s'_{i_0}}$ and $L'_{i_0}$
can be computed likewise.

Let $\Cso =\cso\times S^1$ and $\call_{i_0}=L_{i_0}\times S^1$.
Basically, we compute $\dfp\cap\cso$ and $\dfp\cap L_{i_0}$, as
explained in Section~\ref{Computation_of_df_cap_A}, by computing
$\cup_i(\Hei\cap\Cso)$, $\cup_{i\neq i_0}(\Hei\cap\Cso)$,
$\cup_i(\Hei\cap\call_{i_0})$ and $\cup_{i\neq
i_0}(\Hei\cap\call_{i_0})$.  The properties of the new regions $\Zei
=\Hei\cap\Cso$ and $\Yei =\Hei\cap\call_{i_0}$ are different though
similar to the properties of $\Zsi=\Hsi\cap\Cso$ described in
Section~\ref{Properties_of_zi}.  The analysis of $\Zei$ and $\Yei$ are
subdivided into two parts: first, we consider the line $D_i$
supporting $e_i$ and we examine the regions $\Zdi =\Hdi\cap\Cso$ and
$\Ydi =\Hdi\cap\call_{i_0}$ where $\Hdi$ is the generalized helicoidal
volume induced by $D_i$:
\[\Hdi = \{(P,\theta)\in\Re^2\times S^1\,|\ P\in HD(s,\theta),\ s\in D_i\}.\] 
Then we deduce $\Zei$ (resp. $\Yei$) from $\Zdi$, $\Zsi$ and
$\Z_{s'_i}$ (resp. $\Ydi$, $\Ysi =\Hsi\cap\call_{i_0}$ and $\Ysip$)
where $s_i$ and $s'_i$ are the two endpoints of $e_i$.  Thereafter, we
compute the contribution of \ceio\ to \dfp\ in a way similar to what we did in
Section~\ref{Construction_of_df_cap_cio}.  The following theorem sums
up these results:
\begin{theorem}
\label{thAe} 
We can compute $\dfp\cap \Ae$ and the labels of the edges of \dfp\
incident to the arcs of $\dfp\cap \Ae$ in $O(|\Ae|\alpha_7(n)\log n)$
time using $O(|\Ae|\alpha_8(n))$ space.
\end{theorem}
{
The proof of this theorem, omitted here, is a direct generalization of
the proof of Theorem~\ref{R12}.  Details are given
in~\cite{prisme-3214t} or~\cite{prisme-these-lazard}.
}

\subsection{Arcs of \dfp\ corresponding to the placements where the spider robot is  at the limit of stability}
\label{unstable-equilibrium}

\begin{figure}[th] \begin{center}
  \unitlength 1cm \input{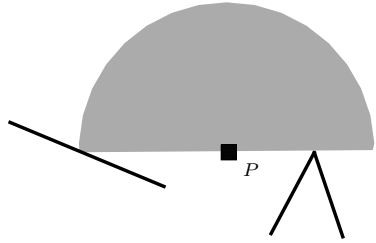} \end{center} \caption{Example of placement $P$ at the limit of stability.}
  \label{limit-stab} \end{figure}

We now have to compute the edges of \Fp\ that do not belong to $\Ae$.
The arcs of $\dfp\cap \Ae$ correspond to placements at the limit of
accessibility of the spider robot, and vice versa. Thus, other edges
of \Fp\ correspond to placements at the limit of stability of the
spider robot. We denote by $\dfp_{stab}$ the set of those edges.  A
placement $P$ of the spider robot is at the limit of stability if and
only if there exists a closed half-disk of radius $R$ centered at $P$
that does not contain any foothold except at least two footholds
located on the diameter of the half-disk such that $P$ is between
these footholds (see Figure~\ref{limit-stab}).  Therefore, the edges
of $\dfp_{stab}$ are portions of the curves drawn by the midpoint of a
ladder of length $2R$ moving by translation and rotation such that the
ladder touches the boundary of the foothold regions in two points but
does not intersect the interior of the foothold regions.  Hence, the
edges of $\dfp_{stab}$ are supported by the projection (onto $\Re^2$)
of the edges of the boundary of the free space of the ladder moving by
translation and rotation amidst the foothold regions considered as
obstacles, \ie, the set of $(P,\theta)\in\Re^2\times
\Re/\pi\ZZ$ such that the ladder of length $2R$ that has its midpoint at
$P$ and makes an angle $\theta$ with the $x$-axis does not 
properly intersect
 the interior of the foothold regions.  According to
~\cite{ss-nempa-87}, the edges of the boundary of the free space of
the ladder can be computed in $O(|\Ae|\log n)$ time using $O(|\Ae|)$
space.  
The projection (onto $\Re^2$) of each edge can easily be computed in constant time.
  Thus, we can compute, in $O(|\Ae|\log n)$ time
and $O(|\Ae|)$ space (using~\cite{ss-nempa-87}), a set of curves in
$\Re^2$ that contains the arcs of \dfp\ that correspond to placements
at the limit of stability of the spider robot.  However, it remains to
compute the portions of these curves that belong to $\dfp$.

\subsubsection{Notations and definitions}

The relative interior of an $e_i$ is called a {\em wall}.  An endpoint
of an $e_i$ is called a {\em corner} (when several walls share an
endpoint, we define only one corner at that point).  
The {\em ladder} is a line segment of  length $2R$.
A {\em placement} of the ladder is a pair $(P,\theta)\in\Re^2\times
\Re/\pi\ZZ$ where $P$ is the location of the midpoint of the ladder
and $\theta$ is the angle between the $x$-axis and the ladder.  A {\em
free placement} of the ladder is a placement where the ladder does not
properly intersect the walls or partially lies on some walls and does
not properly intersect the others (if none of the polygonal regions of
$\cal S$ are degenerated into line segments or points, then a free
placement of the ladder is a placement where the ladder does not
 intersect the interior of the polygonal regions of $\cal S$).  A {\em
placement of type corner-ladder} is a placement of the ladder such
that the relative interior of the ladder touches a {\em corner}.  A
{\em placement of type wall-endpoint} is a placement of the ladder
such that an endpoint of the ladder touches a {\em wall}.  A {\em
placement of type corner-endpoint} is a placement of the ladder such
that an endpoint of the ladder touches a corner.  We now define {\em
$k$-contact placements} of the ladder.
  
 \begin{figure}[th] \begin{center}
  \unitlength 1cm \input{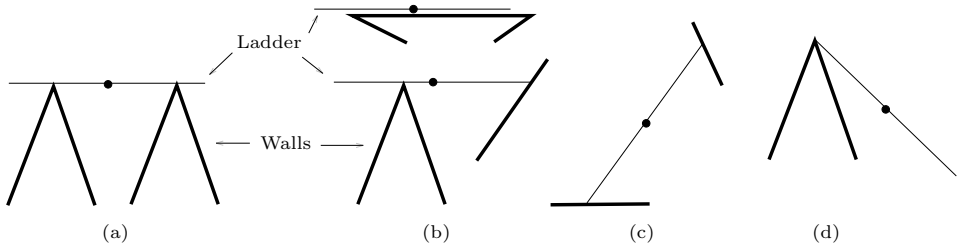} \end{center} \caption{Examples of 2-contact placements of type (a): 
(corner-ladder)$^2$, (b): (corner-ladder, wall-endpoint), (c): (wall-endpoint)$^2$ and (d): (corner-endpoint).}
  \label{Examples-2-contact-placements} \end{figure}

A {\em 1-contact placement} is a free placement of type corner-ladder
or wall-endpoint.  A {\em 2-contact placement} is either the
combination of two 1-contact placements or a free placement of type
corner-endpoint.  A 2-contact placement is said to be of type
(corner-ladder)$^2$, (corner-ladder, wall-endpoint),
(wall-endpoint)$^2$, or (corner-endpoint), in accordance to the types
of placements involved in the 2-contact placement (see
Figure~\ref{Examples-2-contact-placements}).  Given two walls (resp. a
wall and a corner, two corners, one corner), the set of 2-contact
placements induced by these two walls (resp. the wall and the corner,
the two corners, the single corner) is called a {\em 2-contact curve}.
The type of a 2-contact curve is the type of the 2-contact placement
defining the curve.  Note that the 2-contact curves are defined in
$\Re^2\times\Re/\pi\ZZ$.  A {\em 3-contact placement} is a combination
of a 1-contact placement and a 2-contact placement. The types of
3-contact placements are naturally given by (corner-ladder)$^3$,
(corner-endpoint, wall-endpoint)$\ldots$ With this definition, we
unfortunately cannot guarantee that all the 2-contact curves end at
3-contact placements. Indeed, a 2-contact curve defined by the ladder
sliding along a wall (see Figure~\ref{Examples-2-contact-placements}b)
ends on one side (if no other wall blocks the sliding motion) at a
2-contact placement of type (corner-endpoint), where the ladder is
collinear with the wall, without properly intersecting it.  In order
to ensure that all the 2-contact curves end at 3-contact placements,
we consider these 2-contact placements as 3-contact placements, and
denote their type by (corner-endpoint, $\parallel$).  A {\em k-contact
placement}, $k>3$, is the combination of $p$ 1-contact placements, $q$
2-contact placements and $r$ 3-contact placements such that
$p+2q+3r=k$.

Now, we define a {\em 2-contact tracing} as the projection onto
$\Re^2$ of a 2-contact curve.  Similarly as above, we define the types
of the 2-contact tracings.  Notice that, to any point $P$ on a given
2-contact tracing $\calk$, corresponds a unique placement $(P,\theta)$
on the 2-contact curve that projects onto $\calk$.  It follows that,
to any point $P$ on a 2-contact tracing $\calk$, corresponds a unique
pair $(M,N)$ of points of contact between the ladder at $(P,\theta)$
and the walls ($M$ and $N$ are equal when $\calk$ is a 2-contact
tracing of type (corner-endpoint)); when $P$ is an endpoint of
$\calk$, a 3-contact placement corresponds to $P$, however, $(M,N)$ is
uniquely defined by continuity. The points $M$ and $N$ are called the
{\em contact points corresponding to $P\in\calk$.}  We also define the
three contact points corresponding to a 3-contact placement.

A 2-contact tracing is either a straight line segment, an arc of
ellipse, an arc of conchoid or a circular arc.  Indeed (see
Figures~\ref{ce}, \ref{we-we}, \ref{cl-we} and \ref{cl-cl}), a
2-contact tracing of type (corner-endpoint) is a circular arc; a
2-contact tracing of type (wall-endpoint)$^2$ is an arc of ellipse; a
2-contact tracing of type (corner-ladder, wall-endpoint) is an arc of
conchoid (see \cite{prisme-3214t}); a 2-contact tracing of type
(corner-ladder)$^2$ is a straight line segment.  As we said before, we
can compute all these 2-contact tracings in $O(|\Ae|\log n)$ time using
$O(|\Ae|)$ space~\cite{ss-nempa-87}, and it remains to compute the
portions of these curves that belong to $\dfp$.

\subsubsection{Overview}

We first show that only some portions of the 2-contact tracings
correspond to positions at the limit of stability of the spider robot
(Section~\ref{subsubsection-Relevant-2-contact-tracings}). These
portions are called the {relevant} 2-contact tracings.  Then, we prove
that we do not have to take into consideration the intersections
between the relative interior of relevant 2-contact tracings
(Proposition~\ref{A-end-point-of-proper-2-contact-curves}).  We also
show that, if a point $A$ is an endpoint of several relevant 2-contact
tracings, only two of them can support edges of $\dfp_{stab}$ in the
neighborhood of $A$ (Propositions~\ref{at-most-two-edges}).  
Finally (Section~\ref{Construction-of-Delta}), we compute a graph
whose edges are relevant 2-contact tracings and where the degree of
each node is at most two.  This graph induces a set $\Delta$ of curves
supporting $\dfp_{stab}$ (Theorem~\ref{thC}) that will allow us to
compute $\dfp_{stab}$ in Section~\ref{Construction_of_F_2}.

\subsubsection{Relevant 2-contact tracings}
\label{subsubsection-Relevant-2-contact-tracings}

As mentioned above, a placement $P$ of the spider robot is at the
limit of stability if and only if there exists a closed half-disk of
radius $R$ centered at $P$ that does not contain any foothold except
at least two footholds located on the diameter of the half-disk, one
on each side of $P$.  Thus, a point $P$ of a 2-contact tracing $\calk$
belongs to an arc of $\dfp_{stab}$ only if $P$ lies between the two
contact points corresponding to $P\in
\calk$.  The portions of the 2-contact tracings for which that
property holds are called the {\em relevant 2-contact tracings}. The
other portions are called the {\em irrelevant 2-contact tracings}.  We
now show how to compute the relevant 2-contact tracings for each type
of contact.  Let $\calk$ denote a 2-contact tracing, let $P\in\calk$
and let $M$ and $N$ be the two contact points corresponding to
$P\in\calk$. In Figures~\ref{ce}, \ref{we-we},
\ref{cl-we} and \ref{cl-cl}, the walls and the relevant 2-contact
tracings are  thick, the irrelevant 2-contact
tracings are dashed  thick, and the ladder is thin.

\vspace{.5cm}

\vbox{
{\bf Type (corner-endpoint)}:\smallskip\\
\begin{minipage}[b]{7cm}
$\calk$ is a circular arc,  $M$ and $N$ coincide with one endpoint
of the ladder. Thus, all the 2-contact tracings of type
(corner-endpoint) are wholly irrelevant.
\vspace*{1.2cm}
\end{minipage}
\hfill
\begin{minipage}[t]{5.6cm}
\begin{center}
\vspace{-3cm}
\unitlength 0.7cm
\input{ce-v2.ltex}
\end{center}
\refstepcounter{figure}\label{ce}
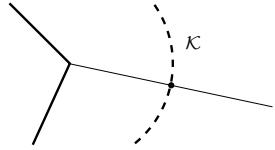
{\footnotesize {\normalfont\scshape \figurename~\thefigure}. {\normalfont\itshape Irrelevant 
2-contact tracing of type (corner-endpoint), \ie, circular arc.}}
\end{minipage}
}

\vbox{
{\bf Type (wall-endpoint)$\mathbf{^2}$}:\smallskip\\
\begin{minipage}[b]{7cm}
$\calk$ is an arc of ellipse, $M$ and $N$ are the endpoints of the
ladder and thus, $P$ lies between them.  Therefore, all the 2-contact
tracings of type (wall-endpoint)$^2$ are wholly relevant.
\vspace*{1cm}
\end{minipage}
\hfill
\begin{minipage}[t]{5.6cm}
\begin{center}
\vspace{-3cm}
\unitlength 0.7cm
\input{we-we-v2.ltex}
\end{center}
\refstepcounter{figure}\label{we-we}
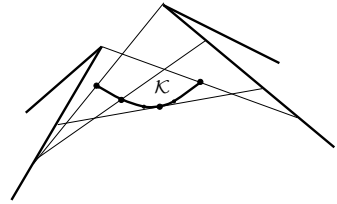
{\footnotesize {\normalfont\scshape \figurename~\thefigure}. {\normalfont\itshape Relevant 
2-contact tracing of type (wall-endpoint)$^2$, \ie, arc of ellipse.}}
\end{minipage}
}

\vbox{
{\bf Type (corner-ladder, wall-endpoint)}:\smallskip\\
\begin{minipage}[b]{7cm}
\vspace*{1cm}
$\calk$ is an arc of conchoid.  If the distance between the corner and
the wall is greater than $R$, then $\calk$ is wholly relevant.
\vspace*{1.3cm}

Otherwise, if that distance is smaller than $R$, then, the two relevant
portions and the irrelevant portion of $\calk$ are incident to the corner
involved in the type of $\calk$.
\vspace*{0.4cm}

Notice that, if the corner is an endpoint of the wall (see
Figure~\ref{Examples-2-contact-placements}b), then $\calk$ degenerates
into a line segment and the irrelevant portion of $\calk$ is the
portion which is not supported by the wall.
\end{minipage}
\hfill
\begin{minipage}[t]{5.6cm}
\begin{center}
\vspace{-7.5cm}
\unitlength 0.7cm
\input{cl-we-v2.ltex}
\refstepcounter{figure}\label{cl-we}
\end{center}
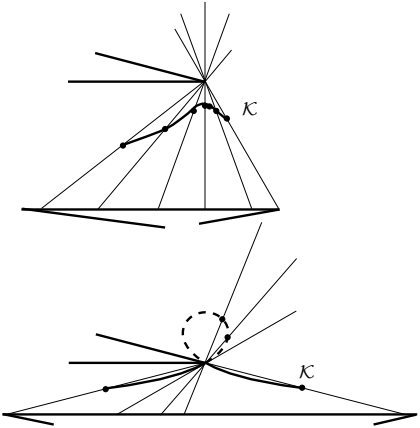
{\footnotesize {\normalfont\scshape \figurename~\thefigure}. {\normalfont\itshape Relevant, 
and  partially relevant, 2-contact tracings of type (corner-ladder, wall-endpoint), \ie, arcs of conchoid.}}
\end{minipage}
\smallskip
}
\smallskip

\vbox{
{\bf Type (corner-ladder)$\mathbf{^2}$}:\smallskip\\
\begin{minipage}[b]{7cm}
$\calk$ is a line segment.  If the distance between the two corners is
greater than $R$, then $\calk$ is wholly relevant; otherwise, the
portion of $\calk$ which is relevant, is the line segment joining the
two corners.
\vspace*{.5cm}
\end{minipage}
\hfill
\begin{minipage}[t]{5.6cm}
\begin{center}
\vspace{-3cm}
\unitlength 0.7cm
\input{cl-cl-v2.ltex}
\refstepcounter{figure}\label{cl-cl}
\end{center}
{\footnotesize {\normalfont\scshape
\figurename~\thefigure}. {\normalfont\itshape Relevant, and partially
relevant, 2-contact tracings of type (corner-ladder)$^2$.}}
\end{minipage}}

We  now show that the intersections between the relative interiors of
the relevant 2-contact tracings are not interesting for the spider robot
motion problem.  We recall that, if a vertex $A$ of \dfp\ belongs to
$\Ae$, then we know by Theorem~\ref{thAe} the labels of the edges of
\dfp\ incident to $A$.  Otherwise, if $A\not\in\Ae$, then the two
edges of \dfp\ that end at $A$ correspond to placements at the limit
of stability of the spider robot.
\begin{proposition}
\label{A-end-point-of-proper-2-contact-curves}
Any vertex $A$ of $\dfp$, such that $A\not\in\Ae$, is an endpoint of the
two relevant 2-contact tracings 
supporting
 the edges of \dfp\ ending at $A$.
\end{proposition}
\proof 
Since the two edges of \dfp\ that end at $A$ correspond to placements
at the limit of stability of the spider robot, they are both supported
by some relevant 2-contact tracings.  Thus, we only have to prove that
$A$ is an endpoint of these two relevant 2-contact tracings.

\begin{figure}[ht] \begin{center}
  \unitlength 1cm \input{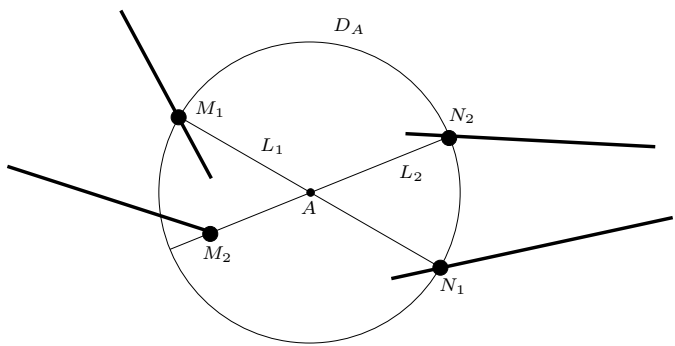} \end{center} \caption{For the proof of Proposition~\ref{A-end-point-of-proper-2-contact-curves}.}
  \label{MMMMeeee} \end{figure}

Let $\calk_1$ and $\calk_2$ be these two relevant 2-contact tracings
and assume for a contradiction that $A$ is not an endpoint of
$\calk_1$ (nothing is assumed for $A$ with respect to $\calk_2$).  Let
$L_1=(A,\theta_1)$ (resp.  $L_2=(A,\theta_2)$) be the placement of the
ladder that correspond to $A\in\calk_1$ (resp. $A\in\calk_2$) and let
$M_1$ and $N_1$ (resp. $M_2$ and $N_2$) be the corresponding contact
points (see Figure~\ref{MMMMeeee}).  First, notice that $L_1\neq
L_2$. Indeed, otherwise, $L_1$ is at least a 3-contact placement and
then, $A$ must be an endpoint of $\calk_1$, which contradicts our
assumption.

By the definition of the relevant 2-contact tracings, $A$ is between
$M_1$ and $N_1$.  Moreover, $A$ cannot be equal to $M_1$ or $N_1$
since $A$ is not an endpoint of $\calk_1$.  It follows that neither
$M_2$ nor $N_2$ is equal to $A$, because otherwise $L_1$ would be a
3-contact placement. Therefore, $A$ is strictly between $M_1$ and
$N_1$, and strictly between $M_2$ and $N_2$.  Thus, $A$ is strictly
inside the polygon $(M_1M_2N_1N_2)$.

On the other hand, since $A\not\in\Ae$, $A$ does not belong to any
$\cei$, and therefore, the walls supporting $M_1$, $N_1$, $M_2$ and $N_2$
intersect the open disk $D_A$ of radius $R$ centered at $A$.  Thus,
there exists four points $M'_1$, $N'_1$, $M'_2$ and $N'_2$ on these
walls and in $D_A$, that are close enough to $M_1$, $N_1$, $M_2$ and
$N_2$ respectively to ensure that $A$ belongs to the interior of the
polygon $(M'_1M'_2N'_1N'_2)$.  Since the distances from $A$ to $M'_1$,
$N'_1$, $M'_1$ and $N'_2$, are strictly smaller than $R$, $A$
belongs to the interior of $\Fp$.  This contradicts our assumption
that $A$ is a vertex of \dfp\ and yields the result.
\endproof

\begin{figure}[th] \begin{center}
  \unitlength 1cm \input{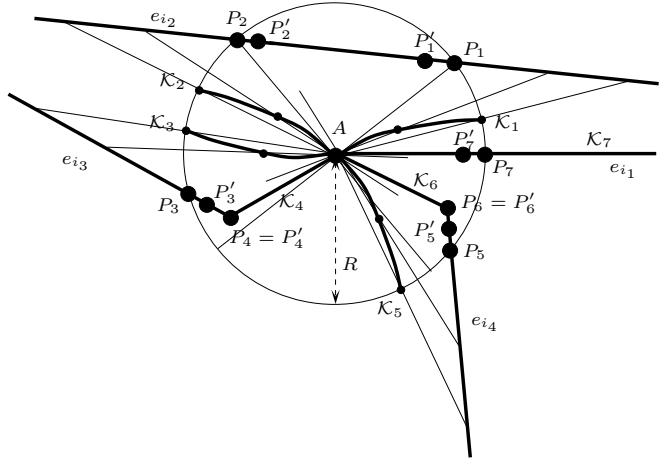} \end{center} \caption{Relevant 2-contact 
tracings $\calk_1,\ldots,\calk_7$ ending at $A$.  $\calk_1, \calk_2,
\calk_3$ and $\calk_5$ are 2-contact tracings of type (corner-ladder,
wall-endpoint) (\ie, arcs of conchoid).  $\calk_7$ is a degenerated
2-contact tracing of type (corner-ladder, wall-endpoint) (\ie, a line
segment).  $\calk_4$ and $\calk_5$ are 2-contact tracings of type
(corner-ladder)$^2$ (\ie, line segments).}
  \label{nodes-of-degree-k_simplifie-v2} \end{figure}

Consider now the adjacency graph $\calg$ of the relevant 2-contact
tracings such that two relevant 2-contact tracings are connected in
$\calg$ if and only if they have a common endpoint (the intersections
between the relative interiors of the relevant 2-contact tracings are
not considered).  Notice that, given the set of relevant 2-contact
tracings, $\calg$ can  be easily computed in $O(|\Ae|\log n)$ time.
Now, given two vertices of $\dfp\cap\Ae$ that are connected along 
\dfp\ by arcs of $\dfp_{stab}$, we want to compute these arcs.
For computing these arcs, we cannot simply use the graph
$\calg$ because the degree of some nodes of $\calg$ may be arbitrarily
large (see Figure~\ref{nodes-of-degree-k_simplifie-v2}).  We show in the
next proposition that we can deduce from $\calg$ a graph $\calg^*$ such
that the degree of each node of $\calg^*$ is at most two and that
$\calg^*$ supports any portion of \dfp\ which is the concatenation of
arcs of $\dfp_{stab}$.

We consider four hypotheses (H1,$\ldots$,H4) that obviate the need to
consider degenerate cases. They are not essential but substantially
simplify the proof of the following proposition.  
The first three hypotheses are made to  ensure
that the degree of each vertex of the free space of the ladder is
three.

\begin{description}
\item[H1] The line segments $e_1,\ldots,e_n$ compose the boundary of a set of non 
degenerated polygons (\ie, no polygon is reduced to a line segment or to a point).
\item[H2] The ladder does not admit any  4-contact placement.
\item[H3] The arc (of conchoid) drawn by an endpoint of the ladder 
when its other endpoint moves along a wall while the ladder remains
in contact with a corner, is not tangent to any other wall.
\item[H4] The ladder does not admit any 3-contact placement when its midpoint is located at a corner.
\end{description}

\begin{proposition} 
\label{at-most-two-edges}
For any node $A$ of $\calg$ of degree $k$ such that $A\not\in\Ae$, at
most two relevant 2-contact tracings can support \dfp\ in a
sufficiently small neighborhood of $A$.  Moreover, we can determine
these at most two curves in $O(k\log k)$ time using $O(k)$ space.
\end{proposition}
\proof 
Let $A\not\in\Ae$ be a node of $\calg$ of degree $k$. We assume that
$k>2$, otherwise Proposition \ref{at-most-two-edges} is trivial.  Let
$\calk_{1},\ldots,\calk_{k}$ be the relevant 2-contact tracings that
end at $A$, and let $L_{i}=(A,\phi_i)$ be the placement of the ladder
that corresponds to $A\in\calk_{i}$.  $D_A$ is the open disk of radius
$R$ centered at $A$.  We distinguish two cases whether $A$ is  a
corner or not.

{\bf Case 1: $A$ is a corner.} 
(See Figure~\ref{nodes-of-degree-k_simplifie-v2}.)

The 2-contact tracing $\calk_{i}$ involves at least another contact
than the corner-ladder contact at $A$. This contact cannot be of type
corner-endpoint by Hypothesis H4. If the contact is of type
wall-endpoint, we define $P_i$ as the contact point between this wall
and the ladder at placement $L_i$ (see Figure~\ref{wedge}).  Since
$A\not\in\Ae$, the wall must intersect $D_A$ and we define $P'_i$ as a
point close to $P_i$ in that intersection.  If the
contact is of type corner-ladder, we define $P_i = P'_i$ as the corner
(distinct from $A$) involved in this contact (notice that $P_i =P'_i\in D_A$ by
Hypothesis {H4}).  

{\bf Fact:} $\mathit{\forall i\neq j,\;\;\phi_i\neq
\phi_j}$.\\ Otherwise, $L_i = L_j$ is a 3-contact placement
contradicting Hypothesis {H4}.

{\bf Fact:} {\em ${A}$ is a non-flat vertex of ${\ch(A,P_1,\ldots,P_k)}$ or
belongs to the interior of~${\Fp}$}.\\ Assume that $A\in\dfp$. Then, $A$
lies on the boundary of $\ch(A,P_1,\ldots,P_k)$ because, otherwise,
the $P'_i$ provide footholds such that the spider robot can move in a
neighborhood of $A$.  Furthermore, $A$ must be a non-flat vertex of
$\ch(A,P_1,\ldots,P_k)$, by Hypothesis {H4}.

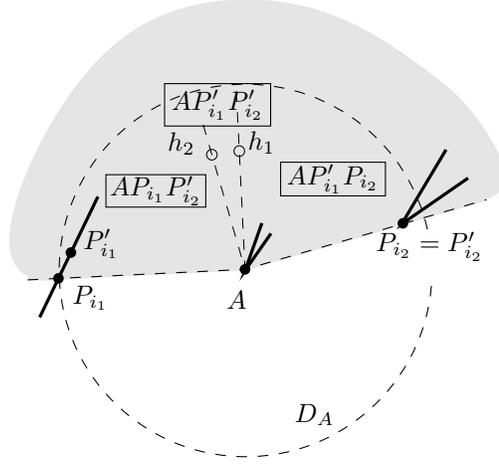
\placeipe{wedge}{Wedge $P_{i_1}AP_{i_2}$  is  in \Fp\ near $A$.}

Assume now that $A\in\dfp$, and let $P_{i_1}$ and $P_{i_2}$ be the two
vertices of $\ch(A,$ $P_1,\ldots,P_k)$ such that $P_{i_1}$, $A$ and
$P_{i_2}$ are consecutive along the boundary of $\ch(A,$
$P_1,\ldots,P_k)$ (see Figure~\ref{wedge}).  We will exhibit a stable
placement for the spider robot at any position $P$ inside the
intersection of the wedge $P_{i_1}AP_{i_2}$ and a neighborhood of $A$.
Let $h_1$ and $h_2$ be two points in the wedge $P_{i_1}AP_{i_2}$ such
that the wedges $P_{i_1}Ah_1$ and $h_2AP_{i_2}$ are right (see
Figure~\ref{wedge}).\\ --- If $P$ is in the wedge $P_{i_1}Ah_2$, and
is close enough to $A$, the footholds $A$, $P_{i_1}$ and $P'_{i_2}$
yield a stable placement for the spider robot.\\ --- If $P$ is in the
wedge $h_2Ah_1$, and is close enough to $A$, footholds $A$, $P'_{i_1}$
and $P'_{i_2}$ yield a stable placement for the spider robot.\\ --- If
$P$ is in the wedge $h_1AP_{i_1}$, and is close enough to $A$,
footholds $A$, $P'_{i_1}$ and $P_{i_2}$ yields a stable placement for
the spider robot.

{\bf Fact:} {\em $\calk_i$, $i\not\in\{i_1,i_2\}$, cannot support an edge of
\dfp\ incident to $A$.}\\
We assume that $A\in\dfp$ because, otherwise, the claim is obvious. 
It follows that $A$ is a non-flat vertex of $\ch(A,P_1,\ldots,P_k)$.
 A 2-contact tracing $\calk_i$, $i\not\in\{i_1,i_2\}$, cannot be an
arc of ellipse because, otherwise, $L_i$ is a 3-contact placement 
(because $A$ is here a corner) contradicting Hypothesis H4.  
Then, $\calk_i$ can be either the segment $AP_i$ or an arc 
of conchoid. If $\calk_i$ is an arc of conchoid, then, by the general
properties of conchoids (see~\cite{prisme-3214t}), $\calk_i$ is
tangent to the segment $AP_i$ at $A$. Thus, $\calk_i$ is always
tangent to the segment $AP_i$ at $A$. 
  The point $P_i$ strictly belongs to the wedge $P_{i_1}AP_{i_2}$,
because we have shown that $\phi_i\not\in\{\phi_{i_1},\phi_{i_2}\}$.
Thus, in a neighborhood of $A$, $\calk_i$ is strictly inside the wedge
$P_{i_1}AP_{i_2}$ and thus
strictly inside $\Fp$.  
Therefore, $\calk_i$ cannot support $\dfp$, in a neighborhood of $A$.

Hence, by sorting the $P_i$ 
by their polar angles
around $A$, we can determine, in
$O(k\log k)$ time, if $A$ is a non-flat vertex of
$\ch(A,P_1,\ldots,P_k)$, and if so, determine $i_1$ and $i_2$. 
If $A$ is a non-flat vertex of $\ch(A,P_1,\ldots,P_k)$,
then, only $\calk_{i_1}$ and $\calk_{i_2}$ can support an edge of
\dfp\ incident to $A$. 
Otherwise, $A$ belongs to the interior of \Fp\ and none of the
2-contact tracings $\calk_1,\ldots,\calk_k$ can support an edge of
\dfp\ incident to $A$.

{\bf Case 2: $A$ is not a corner.}

{\bf Fact:} {\em If there exists $i\neq j$ such that $\phi_i\neq \phi_j$,
then $A$ belongs to the interior of $\Fp$.}\\ For each relevant
2-contact placement $L_i =(A,\phi_i)$, there exists two contact points
$M_i$ and $N_i$ on each side of $A$ at distance less or equal to $R$.
Since $A$ is not a corner, neither $M_i$ nor $N_i$ is equal to $A$,
thus $A$ belongs to the relative interior of the segment $M_iN_i$.  It
follows, when $\phi_i\neq \phi_j$, that $A$ belongs to the interior of
the polygon $(M_iM_jN_iN_j)$ (see Figure~\ref{MMMMeeee}).  Similarly
as in the proof of
Proposition~\ref{A-end-point-of-proper-2-contact-curves}, since
$A\not\in\Ae$, there exists four footholds $M_i',N_i',M_j',N_j'$ in
$D_A$ and in some neighborhoods of $M_i,N_i,M_j,N_j$, respectively,
such that $A$ belongs to the interior of the polygon
$(M_i'M_j'N_i'N_j')$.  Thus, $A$ belongs to the interior of $\Fp$.

Hence, if there exists $i\neq j$ such that $\phi_i\neq \phi_j$, none
of the 2-contact tracings $\calk_1,\ldots,\calk_k$ can support an edge
of \dfp\ incident to $A$.
We now assume that $\phi_i = \phi_j$, $\forall i,j$.

{\bf Fact:} {\em There are at most six 2-contact tracings incident to
$A$}.\\ The general position hypothesis {H2} forbid $k$-contacts
for $k>3$, thus $A$ corresponds to a 3-contact placement.  The three
possible choices of two contacts among three, give three 2-contact
tracing intersecting in $A$ and thus, six arcs incident to $A$.

{\bf Fact:} {\em There are three 2-contact tracings incident to $A$}.\\ If
the 3-contact placement $L$ is of type (corner-endpoint, $\parallel$),
then there are only three 2-contact tracings incident to $A$, that are
two circular arcs and one line segment.  Otherwise, it comes from the
general position hypotheses {H1, H2} and {H3} (designed to ensure that
property) that a 2-contact tracing cannot be valid on both side of the
3-contact, \ie, on one side of the 3-contact placement, the placements
are not free.  The proof that the hypotheses ensured that fact is
detailed in
\cite{prisme-3214t}.

{\bf Fact:} {\em There are two relevant 2-contact tracings incident to $A$}.\\
Since $A$ is not a corner,
 at the 3-contact placement $L$, two contact points are on the same
side of $A$. Thus, only two of the three 2-contact tracings incident
to $A$ are relevant.
\endproof

\subsubsection{Construction of $\Delta$}
\label{Construction-of-Delta}

Now, consider the graph $\calg$ and each node $A$ in turn.  If
$A\in\Ae$, we disconnect all the edges of $\calg$ that end at $A$.
Notice that for each such node $A$, we know, by Theorem~\ref{thAe},
whether $A\in\dfp$ and, in such a case, the labels of the edges of
\dfp\ incident to $A$.  If $A\not\in\Ae$, we disconnect the edges
ending at $A$ except those (at most two) that may support \dfp\ in a
neighborhood of $A$ (see Proposition~\ref{at-most-two-edges}).  In
this way, we obtain a graph $\calg^*$ such that the degree of each
node is one or two.  We consider each connected component of this new
graph as a curve.  Let $\Delta$ be this set of curves. 
These curves are represented in $\calg^*$ as chains (open or closed). 
It follows that, even if a curve is not simple, there exists a natural order 
along the curve. 
Then, according to
Propositions~\ref{A-end-point-of-proper-2-contact-curves} and
\ref{at-most-two-edges}, we get the following theorem:
\begin{theorem}
\label{thC}
We can compute, in $O(|\Ae|\log n)$ time using $O(|\Ae|)$ space, a set
$\Delta$ of curves that support the edges of \dfp\
corresponding to placements at the limit of stability of the spider
robot.  Moreover, any portion $\calp$ of \dfp\ either intersects \Ae\
or belongs to a unique curve of $\Delta$.
\end{theorem}

\subsection{Construction of \Fp\ and \F}
\label{Construction_of_F_2}

We can now construct \Fp\ and $\F$.  Let $\lambda_k(n)$ denote the
maximum length of the Davenport-Schinzel sequence of order $k$ on $n$
symbols and $\alpha_k(n)=\lambda_k(n)/n$. Note that
$\alpha_3(n)=\alpha(n)$.

\begin{theorem} 
\label{thm-tri}
Given, as foothold regions, a set of $n$ non intersecting straight
line segments that satisfies Hypotheses H1, H2, H3 and H4, we can
compute the free space \Fp\ of the spider robot in
$O(|\Ae|\alpha_8(n)\log n)$ time using $O(|\Ae|\alpha_8(n))$ space.
\end{theorem}
\proof  
By Theorem~\ref{thAe}, we can compute the contribution of $\Ae$ to
\dfp\ and the label of the edges of
\dfp\ incident to them in $O(|\Ae|\alpha_7(n)\log n)$
 time using $O(|\Ae|\alpha_8(n))$ space.  By Theorem~\ref{thC}, we can
 compute, in $O(|\Ae|\log n)$ time using $O(|\Ae|)$ space, a set
 $\Delta$ of curves that support the edges of \dfp\ that do not belong
 to $\Ae$.  Moreover, any portion $\calp$ of \dfp\ such that
 $\calp\cap\Ae=\emptyset$ belongs to a unique curve of $\Delta$.
 Thus, by sorting all the vertices of $\dfp\cap\Ae\cap\Delta$ on the
 relevant curves of $\Delta$, we obtain all the edges of \dfp\ that
 belong to a connected component of \dfp\ intersecting $\Ae$.  Indeed,
 for each vertex $A\in\dfp\cap\Ae\cap\Delta$, we know, in a
 neighborhood of $A$, the portion of the curve of $\Delta$ that
 belongs to \dfp\ because we can simply determine, for each edge, a
 side of the edge that belongs to $\Fp$ (the contact points
 corresponding to the edges determine a side that necessarily belongs
 to $\Fp$)\footnote{Observe that when the edge belongs to $\Fp$, its
 two sides belong to $\Fp$.}.  Then, it is an easy task to deduce all
 the connected components of \dfp\ that intersect $\Ae$.

It remains to compute the connected components of \dfp\ that do not
intersect $\Ae$. Each of these components must be a closed curve of
$\Delta$.  Moreover, all the curves of $\Delta$ belong to $\Fp$.
Thus, according to Theorem~\ref{thC}, any closed curve $\calk$ of
$\Delta$ that does not intersect \Ae\ is either a connected component
of \dfp\ or is strictly included in $\Fp$.  Therefore, by considering,
in addition, all the closed curves of $\Delta$ that do not intersect
$\Ae$, we finally obtain a set $\Psi$ of closed curves that contains
\dfp\ and such that any curve of $\Psi$ is either a connected
component of \dfp\ or is strictly included in $\Fp$.

At last, as we can simply determine, for each curve of $\Psi$, a side
of the edge that belongs to $\Fp$, we can easily deduce from $\Psi$ the
free space $\Fp$. That concludes the proof since all these computations
can be done in $O(|\Ae|\alpha_8(n)\log n)$ time using
$O(|\Ae|\alpha_8(n))$ space.
\endproof 

As we said at the beginning of
Section~\ref{polygonal-foothold-regions}, the free space of the spider
robot using as foothold regions a set of polygonal regions is obtained
by adding these polygonal regions to $\Fp$.  This does not increase
the geometric complexity of the free space nor the complexity of the
computation.  Thus, we get the following theorem:
\begin{theorem}
\label{ER} 
Given a set of pairwise disjoint polygonal foothold regions with $n$
edges in total that satisfies Hypotheses H1, H2, H3 and H4, we can
compute the free space \F\ of the spider robot in
$O(|\Ae|\alpha_8(n)\log n)$ time using $O(|\Ae|\alpha_8(n))$ space.
\end{theorem}

The function $\alpha_8(n)$ is extremely slowly growing and can be
considered as a small constant in practical situations.  This result
is almost optimal since, as shown in~\cite{bddp-mplrs-95},
$\Omega(|\Ae|)$ is a lower bound for the size of $\F$.

\section{Conclusion}
\label{conclusion}

We have seen in Theorem~\ref{R} that, when the foothold regions are
$n$ points in the plane, the free space of the spider robot can be
computed in $O(|\A|\log n)$ time using $O(|\A|\alpha(n))$ space where
$\alpha(n)$ is the pseudo inverse of the Ackerman's function and \A\
the arrangement of the $n$ circles of radius $R$ centered at the
footholds.  By~\cite{bddp-mplrs-95} the size of \F\ is known to be
$\Theta(|\A|)$.  The size of \A\ is $O(n^2)$ but it has been shown
in~\cite{s-ksacs-91} that $|\A|=O(kn)$, where $k$ denotes the maximum
number of disks of radius $R$ centered at the footholds that can cover
a point of the plane.  Thus, in case of sparse footholds, the sizes of
\A\ and \F\ are linearly related to the number of footholds. 

When the foothold regions are 
polygons with $n$ edges in total, the free space of the spider robot
can be computed in $O(|\Ae|\alpha_8(n)\log n)$ time using
$O(|\Ae|\alpha_8(n))$ space, where $n\alpha_k(n)=\lambda_k(n)$ is the
maximum length of a Davenport-Schinzel sequence of order $k$ on $n$
symbols, and \Ae\ is the arrangement of the $n$ curves consisting of the
points lying at distance $R$ from the straight line edges.  Note that
the size of \Ae\ is $O(n^2)$.

It should be observed that, in the case of point footholds, our
algorithm implies that $O(|\A|\alpha(n))$ is an upper bound for
$|\F|$. However, this bound is not tight since
$|\F|=\Theta(|\A|)$~\cite{bddp-mplrs-95}. In the case of polygonal
footholds, our analysis implies that $O(|\Ae|\alpha_8(n))$ is an upper
bound for $|\F|$. We leave as an open problem to close the (small) gap
between this upper bound and the $\Omega(|\Ae|)$ lower bound.

Once the free space \F\ is known, several questions can be
answered. In particular, given two points in the same connected
component of $\F$, the algorithm in~\cite{bddp-mplrs-95}
computes a motion of the spider robot, \ie, a motion 
of the body and a corresponding sequence of leg assignments that
allows the robot to move from one point to the other.  

The motion planning problem for other types of legged robots
remains to be studied. The case where all the legs are not attached at
the same point on a polygonal/polyhedral body is particularly
relevant. A spider robot for which all the legs are not of the same
length is also an interesting model.

\section*{Acknowledgments}

We would like to thank Joseph O'Rourke for  helpful comments.


\end{document}